\documentclass[acmsmall,numbers,prologue,table]{acmart}

\usepackage[most]{tcolorbox}

\usepackage{graphicx} 
\usepackage{etoolbox} 
\usepackage{wrapfig}
\usepackage{color}

\newcommand{\BfPara}[1]{\vspace{0.3em}{\noindent\bf#1.}\xspace}

\newcommand{\etal}{{\em et al.}\xspace}

\usepackage[font=footnotesize]{caption}
\usepackage{caption}
\usepackage{pdfpages}
\usepackage[framemethod=TikZ]{mdframed}
\definecolor{boxcolor}{RGB}{240, 248, 255} 

\newmdenv[
  backgroundcolor=blue!25,
  linewidth=0pt,
  innerleftmargin=5pt,
  innerrightmargin=5pt,
  innertopmargin=5pt,
  innerbottommargin=5pt
]{takeawaybox}

\usepackage{tikz} 
\usepackage{pgfplots} 
\pgfplotsset{compat=1.18} 

\usepackage{amsmath}
\usepackage{amsmath,amssymb}
\usepackage{hyperref}
\usepackage{booktabs}
\usepackage{multirow}
\usepackage{pgfplots}
\pgfplotsset{compat=1.18}
\usepackage{caption}
\usepackage{subcaption}
\usepackage{tikz}
\usepackage{array}
\usepackage{url}
\usepackage{pifont}
\usepackage{tabularx}
\usepackage{wasysym}

\newcommand{\takeaway}[1]{\begin{tcolorbox}[
  colback=blue!15!white,
  colbacklower=blue!2!white,
  colframe=blue!70!white,
  title=\textbf{Takeaway},
  fonttitle=\sffamily\bfseries,
  enhanced,
  sharp corners=south,
  boxrule=1pt,  
  top=2pt,
  bottom=2pt,
  left=4pt,
  right=4pt,
  enhanced
]
#1
\end{tcolorbox}}

\newcolumntype{L}[1]{>{\raggedright\arraybackslash}m{#1}}
\newcolumntype{C}[1]{>{\centering\arraybackslash}m{#1}}
\newcolumntype{M}[1]{>{\centering\arraybackslash}m{#1}}
\usepackage{xspace}
\usepackage{multirow}
\usepackage{array}    
\usepackage{xspace}

\begin{document}


\title{Evaluating the Vulnerability of ML-Based Ethereum Phishing Detectors to Single-Feature Adversarial Perturbations}


%

\author{Ahod Alghuried}\email{ah104940@ucf.edu}
\affiliation{\institution{University of Central Florida}\city{Florida}\country{USA}}

\author{Ali Alkinoon}\email{alialkinoon@ucf.edu}
\affiliation{\institution{University of Central Florida}\city{Florida}\country{USA}}

\author{Abdulaziz Alghamdi}\email{abdulaziz.alghamdi@ucf.edu}
\affiliation{\institution{University of Central Florida}\city{Florida}\country{USA}}

\author{Soohyeon Choi}\email{soohyeon.choi@ucf.edu}
\affiliation{\institution{University of Central Florida}\city{Florida}\country{USA}}

\author{Manar Mohaisen}\email{m-mohaisen@neiu.edu}
\affiliation{\institution {Northeastern Illinois University}\city{Chicago}\country{USA}}

\author{David Mohaisen}\email{mohaisen@ucf.edu}\authornote{Corresponding author: David Mohaisen}
\affiliation{\institution{University of Central Florida}\city{Florida}\country{USA}}

\begin{abstract}
This paper explores the vulnerability of machine learning models to simple single-feature adversarial attacks in the context of Ethereum fraudulent transaction detection. Through comprehensive experimentation, we investigate the impact of various adversarial attack strategies on model performance metrics. Our findings, highlighting how prone those techniques are to simple attacks, are alarming, and the inconsistency in the attacks' effect on different algorithms promises ways for attack mitigation. We examine the effectiveness of different mitigation strategies, including adversarial training and enhanced feature selection, in enhancing model robustness and show their effectiveness. 
\end{abstract}


 \maketitle
\section{Introduction}\label{sec:introduction}

Machine learning algorithms are being heavily employed in many applications, and issues concerning their robustness and security are becoming increasingly important and concerning. In particular, adversarial attacks on machine learning algorithms pose a serious risk where small perturbations are added to input data leading to misclassifications that undermine the reliability of these algorithms~\cite{SaadSNKSNM20,SaadNKKNM19}. Understanding the vulnerabilities of these models to adversarial attacks is an essential first step toward the development of robust and dependable machine learning-powered systems, especially in security applications.

One of the applications of machine learning algorithms has been detecting fraud in Ethereum, including phishing and fake initial coin offerings (ICOs). In Ethereum, phishing involves an adversary posing as an honest user to deceive benign users into revealing sensitive information or transferring digital currency through deceptive means such as fake websites or smart contracts. These phishing attempts exploit the decentralized and anonymous nature of the blockchain, making it challenging to trace and recover lost assets once the attack is executed~\cite{YangZTZHH23}. In this context, machine learning classifiers that distinguish between benign and phishing attempts are shown to be effective~\cite{AdeniranAM23}. However, despite such effectiveness, various studies have shown issues with the robustness of these algorithms, including adversarial attacks, where research efforts have introduced sophisticated manipulation (adversarial) attacks that lead to misclassification of phishing attempts, thus fooling benign users ~\cite{GoodfellowSS14,AbusnainaWAWWYM21,AlasmaryAJAANM20,AbusnainaJKNM20,AbusnainaKA0AM19,AbusnainaAAAJNM22,AbusnainaAAAJSN22} (more in section~\ref{sec:related}). 

This study explores how adversarial attacks affect machine learning models used for detecting fraudulent transactions, focusing on Random Forest (RF), Decision Tree (DT), and K-Nearest Neighbors (KNN) classifiers. By applying the Fast Gradient Sign Method (FGSM)~\cite{SzegedyZSBEGF13, AdeniranHD24}, a widely known adversarial attack technique, we examine how these models degrade in performance when exposed to subtle yet deceptive adversarial inputs. Beyond measuring accuracy loss, we also analyze the broader implications of these attacks, particularly their potential to disrupt automated systems that rely on these models. A single misclassification could lead to incorrect decisions, amplifying the overall risk. Our goal is to understand the vulnerabilities of these models to adversarial manipulation and explore strategies to strengthen their defenses, ensuring greater robustness against such attacks.

Recent studies have pointed out that machine learning models can be vulnerable to adversarial attacks, a problem that is not just limited to certain algorithms or use cases but is a widespread issue across the field~\cite{PapernotM0JS16}. Our study builds on this understanding by providing an in-depth analysis of how these attacks affect models designed to detect fraud and offering practical solutions to lessen their impact. Our result indicate that not all algorithms handle such manipulations equally highlighting the critical need to carefully select the models and to design robust defense strategies that are tailored to the specific needs of the application. This research also brings to light the ongoing necessity to continually test and update machine learning models, ensuring they can stand up to the ever evolving threats in dynamic environments.

\section{Literature Review}\label{sec:related}

We review works that focus on detecting and mitigating threats within cryptocurrency networks, particularly focusing on Ethereum and Bitcoin. As shown in Table ~\ref{table:RW}, the review covers key areas, including the detection of malicious transactions, the vulnerabilities of machine learning models to adversarial attacks, and the role of Generative Adversarial Networks (GANs) in enhancing detection capabilities. Additionally, it highlights recent innovations in fraud and phishing detection.

\subsection{Detection and Analysis of Malicious Transactions}

\BfPara{Malicious Transactions in Ethereum} Agarwal~\etal~\cite{AgarwalTS22} analyzed malicious Ethereum transactions, revealing how subtle data manipulations can mislead advanced machine learning models, posing security challenges. Rabieinejad~\etal~\cite{RabieinejadYPD23} leveraged Ethereum data to improve cyber threat detection, stressing the need for representative datasets. Sanjalawe~\etal~\cite{SanjalaweA23} utilized the Benchmark Labeled Transactions Ethereum dataset to detect illicit activities, highlighting the role of feature engineering and semi-supervised learning in improving accuracy.

\BfPara{Entity Classification in Cryptocurrency Networks} Zola~\etal~\cite{ZolaSBGU22} used WalletExplorer data to classify Bitcoin addresses, reducing anonymity and helping to identify illicit entities while highlighting the need for privacy-preserving techniques. Yang~\etal~\cite{YangLLWGZ20} used transaction data to detect Bitcoin spam attacks, demonstrating the effectiveness of GRU-based models.

\BfPara{Detection of Cryptomining Attacks} Mozo~\etal~\cite{MozoPPCT21} applied transaction datasets to detect cryptomining attacks within the Monero network. Their research highlighted that leveraging synthetic network traffic data generated through advanced GAN architectures enables high-precision detection of cryptomining activities, even in privacy-focused networks like Monero.

\subsection{Vulnerabilities of ML Models to Adversarial Attacks}  
As cryptocurrencies like Ethereum gain traction, securing ML models for transaction classification is crucial for detecting fraud and predicting outcomes~\cite{SilvaN20,SinghH19}. However, their vulnerability to adversarial attacks remains a major concern~\cite{SaadD23}. Chen~\etal~\cite{ChenZBH19}, Li~\etal~\cite{LiCH19}, and Oliveira~\etal~\cite{OliveiraVSVBVG21} showed how manipulated inputs deceive ML models, threatening classification accuracy. Narodytska and Kasiviswanathan~\cite{NarodytskaK17} highlighted CNN susceptibility, while Cartella~\etal~\cite{CartellaAFYAE21} optimized fraud detection on imbalanced data, achieving a perfect attack success rate. Bhagoji~\etal~\cite{BhagojiHLS18} introduced Gradient Estimation black-box attacks, achieving near-perfect success on DNNs.

\subsection{GANs Enhanced Detection and Defense Strategies}

\BfPara{Data Augmentation and Synthetic Data Generation}  
Generative Adversarial Networks (GANs) generate realistic synthetic data, addressing labeled data scarcity and enhancing ML model training. They are instrumental in creating Adversarial Examples (AEs) and handling data perturbations in cryptocurrency transactions. Fidalgo~\etal~\cite{FidalgoCP22} leveraged GANs to mitigate class imbalance in Bitcoin, improving model performance. Agarwal~\etal~\cite{AgarwalTS22} used Conditional GANs (CTGAN) to generate adversarial Ethereum data, enhancing model robustness.  

\BfPara{Enhancing Detection Capabilities with GANs}  
Rabieinejad~\etal~\cite{RabieinejadYPD23} employed CTGAN and Wasserstein GANs (WGAN) to augment Ethereum transaction datasets, improving cyber threat detection. Sanjalawe~\etal~\cite{SanjalaweA23} used Semi-Supervised GANs and feature extraction for Ethereum dataset perturbation. Zola~\etal~\cite{ZolaBBGU20,ZolaSBGU22} applied GAN-based augmentation to enrich the underrepresented Bitcoin transaction classes.  

\BfPara{Adversarial Examples for Robust Detection} Yang~\etal~\cite{YangLLWGZ20} utilized WGAN-div to generate adversarial examples for Bitcoin spam detection, ensuring stable training and high-quality synthetic data. Mozo~\etal~\cite{MozoPPCT21} employed WGANs to create synthetic network traffic for Monero cryptomining attack detection, enhancing accuracy with realistic data.

\begin{table}[t]
\centering
\caption{Summary of some of the prior work.}
\label{table:RW}\vspace{-4mm}
\scalebox{0.75}{
\begin{tabular}{lcll}
\hline
\textbf{Paper Title} & \textbf{Year} & \textbf{Adversarial Techniques} & \textbf{Applications in Cryptocurrency} \\
\hline
Li~\etal~\cite{LiCGN18} & 2018 & GANs & Anomaly detection, Secure Water Treatment \\
Qingyu~\etal~\cite{GuoLAHHZZ19} & 2019 & IFCM, AIS, R3 & Fraud detection, TaoBao \\
Ba~\etal~\cite{Ba19} & 2019 & GANs & Credit card fraud, 31-feature dataset \\
Ngo~\etal~\cite{NgoWKPAL19} & 2019 & GANs & Anomaly detection, MNIST, CIFAR10 \\
Zola~\etal~\cite{ZolaBBGU20} & 2020 & GANs, data augmentation & Bitcoin, WalletExplorer (categorized addresses) \\
Yang~\etal~\cite{YangLLWGZ20} & 2020 & WGAN-div, GRU-based detection & Bitcoin, custom spam transaction dataset \\
Shu~\etal~\cite{ShuLKT20} & 2020 & GANs & Intrusion detection, network traffic \\
Mozo~\etal~\cite{MozoPPCT21} & 2021 & WGANs, synthetic traffic & Monero, custom cryptomining dataset \\
Fursov~\etal~\cite{FursovMKKRGBK0B21} & 2021 & Black-box attacks & Transaction records \\
Agarwal~\etal~\cite{AgarwalTS22} & 2022 & CTGAN, K-Means Clustering & Ethereum, Etherscan dataset (2,946 malicious accounts) \\
Fidalgo~\etal~\cite{FidalgoCP22} & 2022 & GANs, data augmentation & Bitcoin, Elliptic dataset (200K+ transactions) \\
Zola~\etal~\cite{ZolaSBGU22} & 2022 & Various GANs, adversarial learning & Bitcoin, WalletExplorer (16M+ addresses) \\
Rabieinejad~\etal~\cite{RabieinejadYPD23} & 2023 & CTGAN, WGAN & Ethereum, 57K normal, 14K abnormal transactions \\
Sanjalawe~\etal~\cite{SanjalaweA23} & 2023 & GANs, feature extraction & Ethereum, labeled transactions (normal, abnormal) \\
\hline
\end{tabular}}
\vspace{-5mm}
\end{table}

\subsection{Fraud and Phishing Detection in Ethereum}
\BfPara{Fraud Detection Techniques} Cartella~\etal~\cite{CartellaAFYAE21} explores how adversarial attacks can be adapted for fraud detection systems dealing with imbalanced data, showing that these attacks can successfully bypass AI models while remaining hard to detect. Ravindranath~\etal~\cite{RavindranathNSBB24} investigates ensemble machine learning models such as CATBoost and LGBM to detect fraud in Ethereum, emphasizing how oversampling techniques improve model accuracy and robustness. Kabla~\etal~\cite{KablaAMACAK22} reviews the applicability of Intrusion Detection Systems (IDS) in detecting attacks on Ethereum-based Decentralized Applications (DApps), discussing vulnerabilities, existing detection methods, and future directions.

\BfPara{Phishing Detection in Ethereum Networks} Tan~\etal~\cite{TanTZZXL23} proposes using Graph Convolutional Networks to detect fraudulent transactions by leveraging network embeddings derived from Ethereum transaction records. Yang ~\etal~\cite{YangZTZHH23} introduces a phishing detection method that enhances interpretability by extracting detailed features from transaction networks, improving detection precision and clarity. Chen~\etal~\cite{ChenLHXZL24} describes a hybrid graph neural network model combined with data augmentation to detect phishing scams on Ethereum, showing superior results by integrating temporal and structural features. Luo~\etal~\cite{LuoQWL24} presents a model using the bias2vec network embedding algorithm to effectively classify Ethereum accounts as phishing or benign.

\BfPara{Graph-Based Phishing Detection Methods} Lv~\etal~\cite{LvD23} proposes a graph-based method using imgraph2vec for phishing detection, showing improved feature extraction and classification performance. Yin~\etal~\cite{YinY23} introduces a community-enhanced graph neural network model that improves phishing detection by analyzing the community structure within Ethereum transaction networks. Wu~\etal~\cite{WuYLYCCZ22} proposes the trans2vec algorithm to detect phishing scams on Ethereum by embedding features from transaction records, demonstrating effectiveness in classification tasks.

\BfPara{Research Gap}  
Despite advancements in machine learning and Despite significant progress in machine learning and blockchain technologies, we still face challenges in defending against adversarial attacks and new threats within cryptocurrency networks. Previous studies have tackled issues like fraud detection and categorizing transactions. However the problem of adversarial examples (AEs) specifically designed to exploit weaknesses unique to cryptocurrencies has not been thoroughly examined. This is especially true for AEs that manipulate features to target transaction fraud, smart contract vulnerabilities, and different types of attacks, which all require more in-depth study.

Our research fills this gap by testing the resilience of machine learning based phishing detection algorithms against {\em simple manipulations}. Focusing on subtle but realistic changes to features to highlight underlying weaknesses in the models and to develop practical, effective countermeasures. Our method involves assessing how vulnerable these systems are to adversarial and identifying which algorithms can best withstand such challenges through a detailed comparative analysis.

\section{Research Questions}\label{RQ}
This research examines how well machine learning algorithms can detect phishing activities in Ethereum transactions, especially when facing adversarial threats. Since securing these transactions is essential, it is important to assess how different models perform under such attacks and explore ways to strengthen their reliability. This study aims to answer key questions that highlight major challenges and emphasize the need to incorporate insights from previous studies into our analysis.

\noindent {\bf RQ1. }{\bf Are machine learning-based phishing detection algorithms robust against simple feature manipulations?}  
ML models are vulnerable to adversarial attacks, where minor input modifications can alter classification outcomes~\cite{NarodytskaK17,CartellaAFYAE21,BhagojiHLS18,SilvaN20,ChenZBH19}. Small changes in transaction features, such as amounts or timestamps, may enable adversaries to evade detection. Evaluating model robustness against such manipulations is essential to ensure reliable Ethereum phishing detection.  

\noindent{\bf RQ2. }{\bf How do different machine learning algorithms compare in robustness against adversarial manipulations in phishing detection?}  
ML algorithms vary in resilience to adversarial attacks~\cite{SinghH19,OliveiraVSVBVG21,FidalgoCP22,AgarwalTS22,RabieinejadYPD23,CroceASDFCM021}. Some models withstand adversarial manipulations better than others~\cite{LiCGN18,DingWZYFG19,Stutz0S19}. A comparative analysis is crucial to identifying the most robust algorithms for securing Ethereum transaction classification.  

\noindent{\bf RQ3.} {\bf How can adversarial manipulations be mitigated in ML-based Ethereum phishing detection?}  
Defensive strategies are needed to counter adversarial attacks~\cite{YangLLWGZ20,MozoPPCT21,CarmonRSDL19,CohenRK19,XieWMYH19}. Mitigation techniques include enhanced feature selection, data augmentation, and adversarial training~\cite{LiCH19,RabieinejadYPD23,ZolaBBGU20}. Strengthening these defenses improves detection accuracy and enhances Ethereum security.

\section{Methodology}\label{sec:methodology}

\begin{figure}[t]
  \hspace{-0.77cm}
  \centering
  \includegraphics[width=10.5cm]{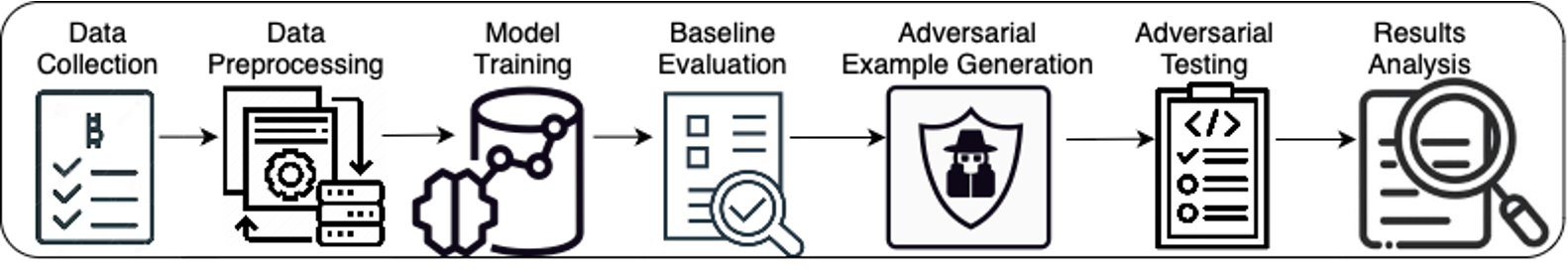}
  \vspace{-2mm}
  \caption{Pipeline in Ethereum transactions and adversarial testing.}
  \label{fig:method}
\end{figure}

The components of our pipeline shown in \autoref{fig:method} are data collection, data preprocessing, modeling training, baseline evaluation, AEs testing, and results and analysis. In the following, we review these components in more detail.

\subsection{Data Preparation}
\subsubsection{Dataset Description}

We used two datasets in our analysis. The first dataset, described by Kabla et al. ~\cite{KablaAMK22}, is implemented for binary classification, phishing and legitimate transaction. It contains different features associated with each transaction, such as \texttt{TxHash}, a unique number that the block assigns to each transaction, and \texttt{BlockHeight}, which indicates the position where the transaction was recorded. The dataset also has \texttt{TimeStamp} which indicates the time when the transaction was approved and written into the blockchain. Inaddition, it contains information about the sender and receiver as the \texttt{From} and \texttt{To} address fields, storing the involved parties' Ethereum addresses. The \texttt{Value} attribute indicates how much Ether was transferred in each transaction, and \texttt{ContractAddress} specifies any smart contract related to the transaction. Moreover, the \texttt{Input} captures any extra data in the transaction. The dataset labels each transaction under the \texttt{Class} feature, marking them either phishing (1) or benign (0). In total, this dataset comprises 23,472 transactions, with 15,989 categorized as benign and 7,483 as phishing.

The second dataset is intended for multi-class classification, categorizing transactions into Phishing, Scam, Fake ICO, or Benign~\cite{Al-EmariASM20}. The features of this dataset include \texttt{hash}, a unique identifier for each transaction, and \texttt{nonce}, a counter ensuring that each transaction is processed only once. The \texttt{transaction\_index} indicates the transaction's position within the block, while \texttt{from\_address} and \texttt{to\_address} denote the blockchain addresses of the sender and receiver, respectively. The \texttt{value} field specifies the amount of cryptocurrency transferred. The dataset also includes \texttt{gas}, representing the gas limit provided for the transaction, and \texttt{gas\_price}, indicating the price per gas unit. The \texttt{input} field contains additional data attached to the transaction. Moreover, \texttt{receipt\_cumulative\_gas\_used} provides the total gas used by all transactions up to and including the current one within the block, and \texttt{receipt\_gas\_used} specifies the gas consumed by this particular transaction. The \texttt{block\_timestamp} and \texttt{block\_number} identify the time and number of the block that includes the transaction, while \texttt{block\_hash} serves as the block's unique identifier. Lastly, the \texttt{from\_scam} and \texttt{to\_scam} fields indicate whether the sender's and receiver's addresses are associated with scams (0 for no, 1 for yes). The dataset also includes categorical data, \texttt{from\_category} and \texttt{to\_category}, which classify the nature of the participants, such as Phishing, Scamming, or Fake ICO. In this dataset, \texttt{Benign} had 57,000 transactions (79.28\%), \texttt{Scamming} had 11,143 transactions (15.51\%), \texttt{Phishing} had 3,106 transactions (4.32\%), and \texttt{Fake ICO} had only 1 transaction. Dataset 2 includes 71,250 transactions.

To ensure the reliability of our datasets, we implemented a comprehensive data preprocessing phase. Maintaining the quality and consistency of data was paramount, so we tackled issues like missing values in critical fields by assigning reasonable default values—for instance, substituting missing entries in the \texttt{to\_address} field with 'Unknown' and filling absent \texttt{input} fields with '0x'. We also addressed outliers by applying normalization techniques, which helped maintain uniformity across the datasets. Additionally, features with high cardinality such as the \texttt{From} and \texttt{To} addresses were encoded to manage their complexity, and temporal features were standardized to Unix epoch time. These preprocessing steps were crucial in preparing the data for the development of robust machine learning models.

After preprocessing, the next step is feature selection, which involves identifying key content-based and network-based features for phishing detection. Content-based features like \texttt{Value}, \texttt{Gas}, and \texttt{Gas\_Price} describe transaction details, while network-based features such as \texttt{From}, \texttt{To}, \texttt{from\_category}, and \texttt{to\_category} capture transaction behavior. These features helped refine the dataset for analysis. Feature engineering was applied to improve data representation. High-cardinality features were encoded, and \texttt{from\_category} and \texttt{to\_category} were merged into \texttt{combined\_category} to summarize transaction types. The \texttt{input} feature, stored as hexadecimal data, was converted into a numerical format for machine learning. These steps ensured that the data was ready for analysis.

\subsection{Experimental Procedures}

We utilized the two datasets to examine how simple AEs impact the classification accuracy of RF, DT, and KNN classifiers.

\BfPara{Minimal Manipulations} AEs were crafted by manipulating specific features. For the first dataset, we conducted the following simple manipulations.

\begin{itemize} 
\item[\ding{172}] {\em Timestamp Manipulation (\textit{TimeStamp})}: We apply predefined intervals to simulate future events, evaluating model robustness against temporal shifts. These intervals introduce different time variations, allowing us to assess how well classifiers maintain accuracy across changing time frames.
\item[\ding{173}]{\em Value Manipulation (\textit{Value})}: Transaction values were adjusted using two methods: adding a fixed percentage uniformly or applying a random percentage change based on the original value. These adjustments tested the models' sensitivity to value changes and their ability to detect phishing activities despite variations.

\item[\ding{174}]{\em Receiver Address Manipulation (\textit{To})}: The receiver's address was randomly reassigned to other addresses in the dataset to simulate phishing transactions with different destinations. This approach tested the classifiers' ability to detect phishing attempts despite changes in transaction recipients.

\item[\ding{175}]{\em Sender Address Manipulation (\textit{From})}: The sender's address was replaced with different addresses to simulate phishing transactions from various sources. This adjustment assessed the classifiers' ability to detect phishing attempts despite changes in the transaction origin.
\end{itemize}

The two datasets were divided into training and testing sets using an 80-20 ratio to ensure a balanced foundation for model training and evaluation. We trained the RF, DT, and KNN classifiers on the training subsets. We assessed their performance using overall accuracy and class-specific accuracy metrics for each label: benign, phishing, and scamming.

We applied both targeted and untargeted adversarial attacks to the second dataset~\cite{Al-EmariASM20}. AEs were created by modifying key transaction features, enabling the simulation of realistic attacks and the identification of potential vulnerabilities.

\BfPara{Untargeted Attacks} The study of untargeted attacks involved generating AEs through broad, random perturbations across the feature space. Initially, all features were manipulated to assess the model’s overall resilience. Results were compiled into detailed tables, highlighting performance across different perturbations and offering a comprehensive view of feature influence on robustness. Next, individual features were targeted for a more precise evaluation. Modifying \textit{from\_address} and \textit{to\_address} introduced unseen addresses, testing the model’s ability to adapt to origin and destination changes. Adjusting \textit{value}, \textit{gas}, and \textit{gas\_price} simulated fluctuations, assessing sensitivity to transaction cost variations. Manipulating \textit{block\_timestamp} and \textit{block\_number} examined the model’s response to sequence and timing changes. Altering \textit{input}, \textit{receipt\_cumulative\_gas\_used}, and \textit{receipt\_gas\_used} evaluated the impact of content modifications.

\BfPara{Targeted Attacks} We conducted targeted attacks focusing on benign, phishing, and scamming scenarios. We employed two methods for generating these targeted attacks: rule-based and gradient-based (using the Fast Gradient Sign Method).

\noindent{\em Rule-based Modifications} This approach adjusts key features, such as transaction value and timestamps, following predefined rules to introduce controlled variations that may influence classification. Additionally, random hexadecimal strings were generated to replace existing \textit{from\_address} and \textit{to\_address}, creating unseen addresses. This process evaluated the model’s ability to recognize phishing attempts when transaction origins and destinations changed and examined its effectiveness in detecting complex fraud attempts.

\begin{itemize}
    \item \textbf{Benign Targeted Attack:} This scenario evaluated the model’s ability to maintain a benign classification despite manipulations, mimicking tactics used to disguise malicious activities. Artificial benign transactions were generated by slightly adjusting features such as \textit{transaction value} and \textit{block\_timestamp}. For instance, minor modifications to transaction values tested whether small financial changes could impact classification accuracy, potentially leading to misclassifications.  

    \item \textbf{Phishing Targeted Attack:}  This attack altered phishing-labeled transactions to make them appear legitimate. Adjustments were made to \textit{from\_address}, \textit{to\_address}, and \textit{value} to mimic attempts to bypass security checks. This method tested whether the model could still identify phishing transactions after changes to key transaction details. 

    \item \textbf{Scamming Targeted Attack:} Scam-labeled transactions were manipulated to test whether they could be misclassified as benign or other category. Adjustments to \textit{gas} and \textit{gas\_price} simulated efforts to obscure scams by modifying transaction costs. This analysis provided insights into the model’s sensitivity to economic factors and its ability to detect scams despite cost-based manipulations.  
\end{itemize}

\paragraph{FGSM}
FGSM introduced small, calculated changes to influence the model’s predictions while keeping feature values within a realistic range. Unlike the rule-based method, it preserved the dataset’s original distribution and adjusted features based on gradients to increase the likelihood of misclassification. This method tested the model’s sensitivity to adversarial modifications. By keeping transactions similar to legitimate ones, FGSM created slight variations that resulted in incorrect classifications. This approach helped identify specific weaknesses in the model’s ability to detect phishing transactions and provided insights into areas for improvement.

\begin{itemize}
    \item \textbf{FGSM Details:} FGSM calculates perturbations that align with the gradient direction of the loss function. The perturbations were applied using the formula:
    \[
    x' = x + \epsilon \cdot \text{sign}(\nabla_x J(\theta, x, y)), 
    \]

    where \( x \) is the original feature, \( x' \) is the perturbed feature (AE), \( \epsilon \) is a small scalar controlling the perturbation's magnitude, \( \nabla_x \) represents the gradient of the loss function \( J \) for \( x \), and \( J(\theta, x, y) \) is the loss function dependent on the model parameters \( \theta \), input \( x \), and true label \( y \). The term \(\text{sign}(\nabla_x J(\theta, x, y))\) provides the direction in which to perturb the feature vector to maximize the loss function.

    \item \textbf{Features:} FGSM modifications were applied to transaction value, gas, gas\_price, and block\_    timestamp. These features play a key role in determining transaction type, and even small changes can influence classification. The method introduced minimal but targeted adjustments, following the gradient direction to increase the likelihood of prediction errors.

\end{itemize}

\BfPara{Evaluation Metrics} 
The performance of the model was assessed using precision, recall, accuracy, and F1-score. The precision is defined as 
\(\text{Precision} = \frac{TP}{TP + FP}\), \(
\text{Recall} = \frac{TP}{TP + FN}\), \(
\text{Accuracy} = \frac{TP + TN}{TP + TN + FP + FN}
\), and \(
\text{F1-score} = 2 \times \frac{\text{Precision} \times \text{Recall}}{\text{Precision} + \text{Recall}}
\) where \( TP \) (True Positives) refers to correctly predicted positive cases, \( TN \) (True Negatives) refers to correctly predicted negative cases, \( FP \) (False Positives) refers to incorrectly predicted positive cases and \( FN \) (False Negatives) refers to incorrectly predicted negative cases.

\section{Results and Analysis}

\subsection{Preliminary Results}
We evaluate how different perturbations on the first dataset affect the accuracy and resilience of RF, DT, and KNN models. We examine how uniform and proportional value manipulations and address change impact model performance. 

\BfPara{Timestamp Manipulations} We evaluated how timestamp changes affected classifier accuracy by applying shifts of +24 hours, +1 hour, +30 minutes, +15 minutes, and +5 minutes. These shifts represented transaction delays or intentional modifications to evade detection.
The RF classifier showed the least impact, with accuracy dropping slightly from 0.98 to 0.95 for a one-day shift and to 0.97 for a one-hour shift. DT experienced a larger decline, with accuracy decreasing from 0.98 to 0.94 under a one-day shift. KNN was the most affected, with accuracy falling from 0.94 to 0.83 for the same change. These results indicate that RF is more resilient to timestamp variations, making it a strong option for phishing detection when transaction timings are altered (Table \ref{table:time_minap}).

\begin{table*}[t]
\centering
\caption{Performance of Random Forest (RF), Decision Tree (DT), and K-Nearest Neighbors (KNN) models under timestamp manipulations across different increments. Metrics include accuracy, precision, recall, F1-score, and count for benign and phishing labels. Increments: original (O), +24 hours (+24), +1 hour (+1), +30 minutes (+30), +15 minutes (+15), and +5 minutes (+5). The baseline dataset represents normal conditions, while the adversarial dataset contains manipulated timestamps.}
\label{table:time_manip}\vspace{-3mm}
\scalebox{0.8}{
\begin{tabular}{l|l|c|cc|cc|cc|cc}
\hline
\textbf{Increment} & \textbf{Model} & \textbf{Accuracy} & \multicolumn{2}{c|}{\textbf{Precision}} & \multicolumn{2}{c|}{\textbf{Recall}} & \multicolumn{2}{c|}{\textbf{F1-score}} & \multicolumn{2}{c}{\textbf{Count}} \\  
\cline{4-11}
& & & Benign & Phishing & Benign & Phishing & Benign & Phishing & Benign & Phishing \\  
\hline
\multicolumn{11}{c}{\textbf{Baseline Dataset}} \\  
\hline
O   & RF  & 0.98 & 1.00 & 0.96 & 0.98 & 1.00 & 0.99 & 0.98 & 15989 & 7483 \\  
    & DT  & 0.98 & 1.00 & 0.95 & 0.98 & 1.00 & 0.99 & 0.97 & 15989 & 7483 \\  
    & KNN &0.94 & 1.00 & 0.85 & 0.92 & 1.00 & 0.96 & 0.92 & 15989 & 7483 \\  
\hline
\multicolumn{11}{c}{\textbf{Adversarial Dataset}} \\  
\hline
+24  & RF  & 0.95 & 0.94 & 0.98 & 0.99 & 0.86 & 0.96 & 0.92 & 15498 & 7974 \\  
     & DT  & 0.94 & 0.95 & 0.94 & 0.97 & 0.88 & 0.96 & 0.91 & 15601 & 7871 \\  
     & KNN & 0.83 & 0.84 & 0.82 & 0.94 & 0.62 & 0.88 & 0.70 & 14376 & 9096 \\  
\hline
+1   & RF  & 0.97 & 0.97 & 0.97 & 0.99 & 0.94 & 0.98 & 0.96 & 15498 & 7974 \\  
     & DT  & 0.96 & 0.97 & 0.95 & 0.98 & 0.94 & 0.97 & 0.94 & 15601 & 7871 \\  
     & KNN & 0.89 & 0.93 & 0.84 & 0.93 & 0.84 & 0.93 & 0.84 & 14376 & 9096 \\  
\hline
+30  & RF  & 0.97 & 0.98 & 0.97 & 0.99 & 0.96 & 0.98 & 0.96 & 15498 & 7974 \\  
     & DT  & 0.96 & 0.97 & 0.95 & 0.98 & 0.94 & 0.97 & 0.94 & 15601 & 7871 \\  
     & KNN & 0.91 & 0.95 & 0.85 & 0.92 & 0.90 & 0.94 & 0.87 & 14376 & 9096 \\  
\hline
+15  & RF  & 0.98 & 0.99 & 0.97 & 0.99 & 0.97 & 0.99 & 0.97 & 15498 & 7974 \\  
     & DT  & 0.96 & 0.98 & 0.95 & 0.98 & 0.95 & 0.98 & 0.95 & 15601 & 7871 \\  
     & KNN & 0.91 & 0.95 & 0.85 & 0.93 & 0.9 & 0.94 & 0.87 & 14376 & 9096 \\  
\hline
+5   & RF  & 0.97 & 0.99 & 0.97 & 0.98 & 0.97 & 0.98 & 0.97 & 15498 & 7974 \\  
     & DT  & 0.97 & 0.98 & 0.95 & 0.98 & 0.97 & 0.98 & 0.96 & 15601 & 7871 \\  
     & KNN & 0.93 & 0.98 & 0.85 & 0.92 & 0.96 & 0.95 & 0.9 & 14376 & 9096 \\  
\hline
\end{tabular}}
\end{table*}

\BfPara{Value Manipulations} When transaction values were changed uniformly, RF and DT accuracy dropped to 0.69, making phishing detection less effective. Attackers who consistently modify transaction values could exploit this weakness. DT’s recall for phishing was especially affected, falling to just 0.01\%.In contrast, proportional changes had little effect, with accuracy and other measures staying nearly the same. This suggests that models adapt better to these variations, which reflect normal transaction patterns. KNN performed consistently under both conditions, showing it was less impacted by value changes (Table \ref{table:value_manipulations}).

\begin{table*}[t]
\centering
\caption{Performance evaluation of Random Forest (RF), Decision Tree (DT), and K-Nearest Neighbors (KNN) models subjected to 1\% uniform (U) and proportional (P) value manipulation strategies compared to the original (O). Metrics include accuracy, precision, recall, F1-score, and count for benign and phishing labels.}
\label{table:value_manipulations}\vspace{-3mm}
\scalebox{0.8}{
\begin{tabular}{l|l|c|cc|cc|cc|cc}
\hline
\textbf{Model} & \textbf{Strategy} & \textbf{Accuracy} & \multicolumn{2}{c|}{\textbf{Precision}} & \multicolumn{2}{c|}{\textbf{Recall}} & \multicolumn{2}{c|}{\textbf{F1-score}} & \multicolumn{2}{c}{\textbf{Count}} \\  
\cline{4-11}
& & & Benign & Phishing & Benign & Phishing & Benign & Phishing & Benign & Phishing \\  
\hline
\multicolumn{11}{c}{\textbf{Random Forest (RF)}} \\  
\hline
O  & Original   & 0.99 & 0.98 & 1.00 & 1.00 & 0.99 & 0.99 & 0.99 & 15803 & 7669 \\  
U  & Uniform    & 0.69 & 0.96 & 0.68 & 0.02 & 1.00 & 0.03 & 0.81 & 23353 & 119 \\  
P  & Proportional & 0.99 & 0.98 & 1.00 & 1.00 & 0.99 & 0.99 & 0.99 & 15813 & 7659 \\  
\hline
\multicolumn{11}{c}{\textbf{Decision Tree (DT)}} \\  
\hline
O  & Original   &  0.98 & 0.95 & 1.00 & 1.00 & 0.98 & 0.97 & 0.99 & 15601 & 7871 \\  
U  & Uniform    &  0.69 & 0.75 & 0.68 & 0.02 & 1.00 & 0.03 & 0.81 & 23294 & 178 \\  
P  & Proportional & 0.98 & 0.95 & 1.00 & 1.00 & 0.98 & 0.97 & 0.99 & 15619 & 7853 \\  
\hline
\multicolumn{11}{c}{\textbf{K-Nearest Neighbors (KNN)}} \\  
\hline
O  & Original   & 0.96 & 0.89 & 0.99 & 0.98 & 0.94 & 0.93 & 0.97 & 15192 & 8280 \\  
U  & Uniform    & 0.96 & 0.89 & 0.99 & 0.98 & 0.94 & 0.93 & 0.97 & 15193 & 8279 \\  
P  & Proportional & 0.96 & 0.89 & 0.99 & 0.98 & 0.94 & 0.93 & 0.97 & 15192 & 8280 \\  
\hline
\end{tabular}}
\end{table*}

\BfPara{Address Manipulations} The models were tested against AEs by modifying the \texttt{From} and \texttt{To} address features in 5,000, 10,000, and 23,472 transactions. Address changes are common in phishing attacks, where attackers alter source or destination addresses to hide fraudulent transactions. For the RF model, accuracy dropped to 0.87 when the \texttt{From} address was changed (Table \ref{table:from_result}) and to 0.84 for the \texttt{To} address (Table \ref{table:to_result}). Precision and recall for phishing transactions also declined, leading to a lower F1 score. While RF remained stable, these results show that address changes can still affect its performance, which could be a concern in real world. The DT model experienced a smaller accuracy reduction, reaching 0.92 for \texttt{From} address changes (Table \ref{table:from_result}) and 0.93 for \texttt{To} (Table \ref{table:to_result}). However, DT’s ability to handle address modifications better may be due to its decision-making process, which can manage categorical data more effectively. KNN was the most affected, with accuracy falling to 0.85 for \texttt{From} and 0.93 for \texttt{To} changes. This led to sharp declines in precision and recall for phishing transactions (Tables \ref{table:from_result} and \ref{table:to_result}). KNN’s high sensitivity to address modifications suggests it may struggle in environments where attackers frequently change transaction addresses, making it less reliable in detecting phishing.

RF and DT handled timestamp and value changes better than KNN. To make ML models more secure and reliable, it is important to refine their design and include adversarial training. The differences in model performance show why thorough testing is necessary. Next, the analysis will examine the second dataset from Al-Emari~\etal~\cite{Al-EmariASM20}, which is used for multi-class classification.

\begin{table*}[t]
\centering
\caption{Performance evaluation of Random Forest (RF), Decision Tree (DT), and K-Nearest Neighbors (KNN) models under manipulations of the \texttt{From} feature in Ethereum transaction datasets. Metrics include accuracy, precision, recall, F1-score, and count for benign and phishing labels. The baseline dataset (Ba) represents normal conditions, while other values correspond to adversarial manipulations.}
\label{table:from_result}\vspace{-3mm}
\scalebox{0.8}{
\begin{tabular}{l|l|c|cc|cc|cc|cc}
\hline
\textbf{Model} & \textbf{Manipulation} & \textbf{Accuracy} & \multicolumn{2}{c|}{\textbf{Precision}} & \multicolumn{2}{c|}{\textbf{Recall}} & \multicolumn{2}{c|}{\textbf{F1-score}} & \multicolumn{2}{c}{\textbf{Count}} \\  
\cline{4-11}
& & & Benign & Phishing & Benign & Phishing & Benign & Phishing & Benign & Phishing \\  
\hline
\multicolumn{11}{c}{\textbf{Random Forest (RF)}} \\  
\hline
---     & Baseline   &  0.99 & 1.00 & 0.96 & 0.98 & 1.00 & 0.99 & 0.98 & 15708 & 7764 \\  
5000   & Adversarial & 0.96 & 0.96 & 0.97 & 0.98 & 0.91 & 0.97 & 0.94 & 16386 & 7086 \\  
10000  & Adversarial & 0.94 & 0.93 & 0.97 & 0.99 & 0.83 & 0.96 & 0.89 & 17027 & 6445 \\  
23472  & Adversarial &  0.87 & 0.84 & 0.97 & 0.99 & 0.60 & 0.91 & 0.74 & 18877 & 4595 \\  
\hline
\multicolumn{11}{c}{\textbf{Decision Tree (DT)}} \\  
\hline
---     & Baseline   &  0.98 & 1.00 & 0.95 & 0.98 & 1.00 & 0.99 & 0.97 & 15601 & 7871 \\  
5000   & Adversarial & 0.97 & 0.98 & 0.95 & 0.98 & 0.96 & 0.98 & 0.96 & 15935 & 7537 \\  
10000  & Adversarial & 0.96 & 0.96 & 0.95 & 0.98 & 0.91 & 0.97 & 0.93 & 16272 & 7200 \\  
23472  & Adversarial &  0.92 & 0.91 & 0.94 & 0.98 & 0.79 & 0.94 & 0.86 & 17207 & 6265 \\  
\hline
\multicolumn{11}{c}{\textbf{K-Nearest Neighbors (KNN)}} \\  
\hline
---     & Baseline   & 0.94 & 1.00 & 0.85 & 0.92 & 1.00 & 0.96 & 0.92 & 14706 & 8766 \\  
5000   & Adversarial & 0.93 & 0.97 & 0.85 & 0.92 & 0.93 & 0.94 & 0.89 & 15209 & 8263 \\  
10000  & Adversarial & 0.91 & 0.94 & 0.84 & 0.92 & 0.87 & 0.93 & 0.85 & 15767 & 7705 \\  
23472  & Adversarial &  0.85 & 0.86 & 0.81 & 0.93 & 0.67 & 0.89 & 0.74 & 17288 & 6184 \\  
\hline
\end{tabular}}
\end{table*}

\begin{table*}[t]
\centering
\caption{Performance evaluation of Random Forest (RF), Decision Tree (DT), and K-Nearest Neighbors (KNN) models under manipulations of the \texttt{To} feature in Ethereum transaction datasets. Metrics include accuracy, precision, recall, F1-score, and count for benign and phishing labels. The baseline dataset (Ba) represents normal conditions, while other values correspond to adversarial manipulations.}
\label{table:to_result}\vspace{-3mm}
\scalebox{0.8}{
\begin{tabular}{l|l|c|cc|cc|cc|cc}
\hline
\textbf{Model} & \textbf{Manipulation} & \textbf{Accuracy} & \multicolumn{2}{c|}{\textbf{Precision}} & \multicolumn{2}{c|}{\textbf{Recall}} & \multicolumn{2}{c|}{\textbf{F1-score}} & \multicolumn{2}{c}{\textbf{Count}} \\  
\cline{4-11}
& & & Benign & Phishing & Benign & Phishing & Benign & Phishing & Benign & Phishing \\  
\hline
\multicolumn{11}{c}{\textbf{Random Forest (RF)}} \\  
\hline
---     & Baseline   &  0.99 & 1.00 & 0.96 & 0.98 & 1.00 & 0.99 & 0.98 & 15708 & 7764 \\  
5000   & Adversarial & 0.96 & 0.95 & 0.97 & 0.98 & 0.89 & 0.97 & 0.93 & 16558 & 6914 \\  
10000  & Adversarial & 0.92 & 0.91 & 0.97 & 0.99 & 0.79 & 0.95 & 0.87 & 17377 & 6095 \\  
23472  & Adversarial &  0.84 & 0.81 & 0.97 & 0.99 & 0.51 & 0.89 & 0.67 & 19537 & 3935 \\  
\hline
\multicolumn{11}{c}{\textbf{Decision Tree (DT)}} \\  
\hline
---     & Baseline   &  0.98 & 1.00 & 0.95 & 0.98 & 1.00 & 0.99 & 0.97 & 15601 & 7871 \\  
5000   & Adversarial & 0.97 & 0.98 & 0.95 & 0.98 & 0.96 & 0.98 & 0.96 & 15884 & 7588 \\  
10000  & Adversarial & 0.96 & 0.97 & 0.95 & 0.98 & 0.93 & 0.97 & 0.94 & 16142 & 7330 \\  
23472  & Adversarial &  0.93 & 0.92 & 0.94 & 0.98 & 0.83 & 0.95 & 0.88 & 16872 & 6600 \\  
\hline
\multicolumn{11}{c}{\textbf{K-Nearest Neighbors (KNN)}} \\  
\hline
---     & Baseline   &  0.94 & 1.00 & 0.85 & 0.92 & 1.00 & 0.96 & 0.92 & 14706 & 8766 \\  
5000   & Adversarial & 0.94 & 0.99 & 0.85 & 0.92 & 0.98 & 0.95 & 0.91 & 14849 & 8623 \\  
10000  & Adversarial & 0.94 & 0.98 & 0.85 & 0.92 & 0.97 & 0.95 & 0.91 & 14988 & 8484 \\  
23472  & Adversarial &  0.93 & 0.96 & 0.85 & 0.93 & 0.93 & 0.94 & 0.89 & 15351 & 8121 \\  
\hline
\end{tabular}}
\end{table*}

\begin{table*}[t]
\centering
\caption{Impact of adversarial attacks on benign, phishing, and scamming detection for Random Forest (RF), Decision Tree (DT), and K-Nearest Neighbors (KNN). The table presents accuracy changes before and after attacks, instance count shifts, and label changes for phishing (P), benign (B), scamming (S), and fake ICO (F).}
\label{table:Benign-target}\label{table:Phishing-target}\label{table:Scamming-target} 
\vspace{-3mm}
\scalebox{0.85}{
\begin{tabular}{l|cc|cc|cc}
\hline
\multirow{2}{*}{\textbf{Metric}} & \multicolumn{2}{c|}{\textbf{Random Forest (RF)}} & \multicolumn{2}{c|}{\textbf{Decision Tree (DT)}} & \multicolumn{2}{c}{\textbf{K-Nearest Neighbors (KNN)}} \\  
\cline{2-7}
 & \textbf{Pre} & \textbf{Post} & \textbf{Pre} & \textbf{Post} & \textbf{Pre} & \textbf{Post} \\  
\hline
\multicolumn{7}{c}{\textbf{Accuracy Changes Due to Attacks}} \\  
\hline
Benign Detection   & 1.00  & 0.84  & 0.99  & 0.84  & 0.97  & 0.90  \\  
Phishing Detection & 0.96  & 0.01  & 0.96  & 0.01  & 0.41  & 0.02  \\  
Scamming Detection & 0.99  & 0.14  & 0.98  & 0.14  & 0.67  & 0.07  \\  
\hline
\multicolumn{7}{c}{\textbf{Instance Counts Before and After Attacks}} \\  
\hline
Benign   & 11,431  & 12,026  & 11,431  & 12,063  & 11,431  & 12,861  \\  
Phishing & 629     & 187     & 629     & 168     & 629     & 306     \\  
Scamming & 2,189   & 2,033   & 2,189   & 2,018   & 2,189   & 1,083   \\  
\hline
\multicolumn{7}{c}{\textbf{Label Changes Due to Attack}} \\  
\hline
Benign → [P, S, F]      & \multicolumn{2}{c|}{[0, 5, 0]}     & \multicolumn{2}{c|}{[0, 5, 0]}     & \multicolumn{2}{c}{[60, 97, 0]}   \\   
Phishing → [B, S, F]      & \multicolumn{2}{c|}{[325, 117, 0]}  & \multicolumn{2}{c|}{[325, 146, 0]}   & \multicolumn{2}{c}{[372,27, 0]}  \\  
Scamming → [B, P, F]      & \multicolumn{2}{c|}{[274, 0, 0]}   & \multicolumn{2}{c|}{[311, 10, 0]}   & \multicolumn{2}{c}{[1,214, 16, 0]} \\  
\hline
\end{tabular}}
\end{table*}

\subsection{Results of Targeted Attacks}\label{sec:targeted}

\BfPara{Rule-based Modifications}This study first examines rule-based adversarial modifications that target specific transaction features to assess how simple changes can affect model performance. These modifications reflect real-world cases where attackers alter some features to avoid detection. Table \ref{table:Benign-target} shows how these changes impact model performance under adversarial attacks.

{\noindent \ding{172} \em Benign Class.}
RF and DT originally classified benign transactions with high accuracy on the test set. However, when exposed to adversarial changes, their accuracy dropped to 0.84, as shown in Table \ref{table:Benign-target}, a decline of over 0.15. This suggests that even small changes can make it harder for the models to correctly classify transactions, creating a risk where attackers can make fraudulent activities appear legitimate. In contrast, KNN performed better under these conditions, maintaining an accuracy of 0.90. Its method, which groups transactions based on similarity, may help it resist some of these changes. The number of benign transactions affected by adversarial modifications increased: from 11,431 to 12,026 for RF, 12,063 for DT, and 12,861 for KNN, as shown in Table \ref{table:Benign-target}.

{\noindent \ding{173} \em Phishing Class.} RF and DT initially had a phishing detection accuracy of 0.96, as shown in Table \ref{table:Phishing-target}. However, after adversarial changes, their accuracy dropped to 0.01, a decrease of more than 0.95. This shows that both models struggle to detect phishing when key transaction details are modified, making them vulnerable to targeted attacks. KNN, which started with a lower accuracy of 0.41, dropped to 0.02. Even though it had a weaker starting point, its near-total failure under attack highlights how difficult it is to maintain phishing detection when attackers exploit specific weaknesses (see Table \ref{table:Phishing-target}). The number of correctly identified phishing transactions dropped sharply: from 629 to 187 for RF, 168 for DT, and 306 for KNN. Meanwhile, more phishing transactions were misclassified as benign, increasing to 325 for RF and DT and 372 for KNN, as shown in Table \ref{table:Phishing-target}.


{\noindent \ding{174} \em Scamming Class}
RF and DT originally had high accuracy in detecting scamming transactions, at 0.99 and 0.98, as shown in Table \ref{table:Scamming-target}. After adversarial changes, their accuracy dropped to 0.14, a decrease of more than 0.85. This shows that both models are highly affected by small modifications, which could weaken their ability to detect fraud. KNN started with an accuracy of 0.67 but fell to 0.07. While it performed well in general classification, it struggled with scam detection, making it less reliable in such cases. The number of correctly classified scamming transactions dropped from 2,189 to 2,033 for RF, 2,018 for DT, and 1,083 for KNN. Misclassifications as benign were highest for KNN, with 1,214 cases, compared to 274 for RF and 311 for DT, as shown in Table \ref{table:Scamming-target}.

\takeaway{Targeted adversarial attacks undermine the effectiveness of machine learning models in detecting phishing and scamming activities, revealing a critical weakness in their robustness across different algorithms.}

\subsection{Gradient-based Approach Using FGSM}

\begin{figure*}
   \vspace{0.2cm}
   \scalebox{0.88}{
   \begin{tikzpicture}

       \node at (-1.2, 1.5) [xshift=2.4cm]  {\textbf{Random Forest }};

       \begin{axis}[
          at={(-1cm,-1cm)},  
    anchor=south west, 
           ybar,
           ymin = 0, ymax = 1,
           ytick distance = 0.2,
           enlargelimits=0.15,
           width = 0.22\textwidth,
           height = 0.22\textwidth,
           ylabel style ={font = \fontsize{9}{9}\selectfont},
           yticklabel style={font = \fontsize{9}{9}\selectfont},
            symbolic x coords={B, Ph, Sc},
           xtick=data,
            xtick=data,
           bar width=4pt,
           grid=both,grid style={line width=0.1pt, draw=gray!10},major grid style={line width=.1pt,draw=black!30, dashed},
           title={Precision},
           legend style={at={(4.9, 2.10)}, anchor=south, legend columns=2, font=\fontsize{8}{10}\selectfont, /tikz/every even column/.append style={column sep=0.5cm}}, 
           legend cell align={left}
         ]       
           \addplot[ybar, fill=blue!50!white, draw=blue!50!black] coordinates {(B,1.00) (Ph,0.98) (Sc,0.99)};
           \addlegendentry{Baseline};
           \addplot[ybar, fill=red!50!white, draw=red!50!black] coordinates {(B,0.95) (Ph,0.87) (Sc,0.94)};
           \addlegendentry{Adversarial};
       \end{axis}

\begin{axis}[
    at={(0.6cm,-1cm)}, anchor=south west, 
    ybar,
    ymin = 0, ymax = 1,
    ytick distance = 0.2,
    enlargelimits=0.15,
 width = 0.22\textwidth,
    height = 0.22\textwidth,
    ylabel style ={font = \fontsize{9}{9}\selectfont},
    yticklabel style={font = \fontsize{9}{9}\selectfont},
    yticklabels={}, 
    symbolic x coords={B, Ph, Sc},
    xtick=data,
    bar width=4pt,
    xticklabel style={font = \fontsize{9}{9}\selectfont},
    grid=both,grid style={line width=0.1pt, draw=gray!10},major grid style={line width=0.1pt,draw=black!30, dashed},
    title={Recall}
     ]       
           \addplot[ybar, fill=blue!50!white, draw=blue!50!black] coordinates {(B,1.00) (Ph,0.96) (Sc,0.99)};
           \addplot[ybar, fill=red!50!white, draw=red!50!black] coordinates {(B,1.00) (Ph,0.32) (Sc,0.85)};
       \end{axis}

\begin{axis}[
    at={(2.2cm,-1cm)}, anchor=south west,
    ybar,
    ymin = 0, ymax = 1,
    ytick distance = 0.2,
    enlargelimits=0.15,
 width = 0.22\textwidth,
    height = 0.22\textwidth,
    ylabel style ={font = \fontsize{9}{9}\selectfont},
    yticklabel style={font = \fontsize{9}{9}\selectfont},
    yticklabels={}, 
    symbolic x coords={B, Ph, Sc},
    xtick=data,
    bar width=4pt,
    xticklabel style={font = \fontsize{9}{9}\selectfont},
    grid=both,grid style={line width=0.1pt, draw=gray!10},major grid style={line width=0.1pt,draw=black!30, dashed},
    title={F1-Score}
]       
   \addplot[ybar, fill=blue!50!white, draw=blue!50!black] coordinates {(B,1.00) (Ph,0.97) (Sc,0.99)};
           \addplot[ybar, fill=red!50!white, draw=red!50!black] coordinates {(B,0.97) (Ph,0.46) (Sc,0.89)};
       \end{axis}


       

       
       \node at (-2, 1.5) [xshift=8.55cm]  {\textbf{Decision Tree }};

           \begin{axis}[
    at={(4cm,-1cm)}, anchor=south west,
           ybar,
           ymin = 0, ymax = 1,
           ytick distance = 0.2,
           enlargelimits=0.15,
 width = 0.22\textwidth,
    height = 0.22\textwidth,
    ylabel style ={font = \fontsize{9}{9}\selectfont},
    yticklabel style={font = \fontsize{9}{9}\selectfont},
    yticklabels={}, 
    symbolic x coords={B, Ph, Sc},
    xtick=data,
    bar width=4pt,
    xticklabel style={font = \fontsize{9}{9}\selectfont},
    grid=both,grid style={line width=0.1pt, draw=gray!10},major grid style={line width=0.1pt,draw=black!30, dashed},
    title={Precision}
                 ]       
         ] 
         
           \addplot[ybar, fill=blue!50!white, draw=blue!50!black] coordinates {(B,1.00) (Ph,0.95) (Sc,0.99)};
           \addplot[ybar, fill=red!50!white, draw=red!50!black] coordinates {(B,1.00) (Ph,0.04) (Sc,0.19)};
       \end{axis}

    \begin{axis}[
    at={(5.6cm,-1cm)}, anchor=south west,
    ybar,
    ymin = 0, ymax = 1,
    ytick distance = 0.2,
    enlargelimits=0.15,
 width = 0.22\textwidth,
    height = 0.22\textwidth,
    ylabel style ={font = \fontsize{9}{9}\selectfont},
    yticklabel style={font = \fontsize{9}{9}\selectfont},
    yticklabels={}, 
    symbolic x coords={B, Ph, Sc},
    xtick=data,
    bar width=4pt,
    xticklabel style={font = \fontsize{9}{9}\selectfont},
    grid=both,grid style={line width=0.1pt, draw=gray!10},major grid style={line width=0.1pt,draw=black!30, dashed},
    title={Recall}
      ]       
         ]       
           \addplot[ybar, fill=blue!50!white, draw=blue!50!black] coordinates {(B,1.00) (Ph,0.97) (Sc,0.99)};
           \addplot[ybar, fill=red!50!white, draw=red!50!black] coordinates {(B,0.02) (Ph,0.72) (Sc,0.29)};
       \end{axis}

   \begin{axis}[
    at={(7.2cm,-1cm)}, anchor=south west,
    ybar,
    ymin = 0, ymax = 1,
    ytick distance = 0.2,
    enlargelimits=0.15,
 width = 0.22\textwidth,
    height = 0.22\textwidth,
    ylabel style ={font = \fontsize{9}{9}\selectfont},
    yticklabel style={font = \fontsize{9}{9}\selectfont},
    yticklabels={}, 
    symbolic x coords={B, Ph, Sc},
    xtick=data,
    bar width=4pt,
    xticklabel style={font = \fontsize{9}{9}\selectfont},
    grid=both,grid style={line width=0.1pt, draw=gray!10},major grid style={line width=0.1pt,draw=black!30, dashed},
    title={F1-Score} 
      ]       
         ]        
           \addplot[ybar, fill=blue!50!white, draw=blue!50!black] coordinates {(B,1.00) (Ph,0.96) (Sc,0.99)};
           \addplot[ybar, fill=red!50!white, draw=red!50!black] coordinates {(B,0.05) (Ph,0.08) (Sc,0.23)};
\end{axis}


        \node at (-2, 1.5) [xshift=13.15cm]  {\textbf{K-Nearest Neighbors}};
\begin{axis}[
    at={(9.0cm,-1cm)}, anchor=south west,
    ybar,
    ymin = 0, ymax = 1,
    ytick distance = 0.2,
    enlargelimits=0.15,
 width = 0.22\textwidth,
    height = 0.22\textwidth,
    ylabel style ={font = \fontsize{9}{9}\selectfont},
    yticklabel style={font = \fontsize{9}{9}\selectfont},
    yticklabels={}, 
    symbolic x coords={B, Ph, Sc},
    xtick=data,
    bar width=4pt,
    xticklabel style={font = \fontsize{9}{9}\selectfont},
    grid=both,grid style={line width=0.1pt, draw=gray!10},major grid style={line width=0.1pt,draw=black!30, dashed},
    title={Precision} 
      ]       
         ]               
    \addplot[ybar, fill=blue!50!white, draw=blue!50!black] coordinates {(B,0.92) (Ph,0.78) (Sc,0.82)};
    \addplot[ybar, fill=red!50!white, draw=red!50!black] coordinates {(B,0.80) (Ph,0.00) (Sc,0.00)};
\end{axis}

\begin{axis}[
    at={(10.6cm,-1cm)}, anchor=south west,
    ybar,
    ymin = 0, ymax = 1,
    ytick distance = 0.2,
    enlargelimits=0.15,
 width = 0.22\textwidth,
    height = 0.22\textwidth,
    ylabel style ={font = \fontsize{9}{9}\selectfont},
    yticklabel style={font = \fontsize{9}{9}\selectfont},
    yticklabels={}, 
    symbolic x coords={B, Ph, Sc},
    xtick=data,
    bar width=4pt,
    xticklabel style={font = \fontsize{9}{9}\selectfont},
    grid=both,grid style={line width=0.1pt, draw=gray!10},major grid style={line width=0.1pt,draw=black!30, dashed},
    title={Recall} 
      ] 
]       
    \addplot[ybar, fill=blue!50!white, draw=blue!50!black] coordinates {(B,0.97) (Ph,0.41) (Sc,0.67)};
    \addplot[ybar, fill=red!50!white, draw=red!50!black] coordinates {(B,1.00) (Ph,0.00) (Sc,0.00)};
\end{axis}

\begin{axis}[
     at={(12.2cm,-1cm)}, anchor=south west,
    ybar,
    ymin = 0, ymax = 1,
    ytick distance = 0.2,
    enlargelimits=0.15,
 width = 0.22\textwidth,
    height = 0.22\textwidth,
    ylabel style ={font = \fontsize{9}{9}\selectfont},
    yticklabel style={font = \fontsize{9}{9}\selectfont},
    yticklabels={}, 
    symbolic x coords={B, Ph, Sc},
    xtick=data,
    bar width=4pt,
    xticklabel style={font = \fontsize{9}{9}\selectfont},
    grid=both,grid style={line width=0.1pt, draw=gray!10},major grid style={line width=0.1pt,draw=black!30, dashed},
    title={F1-Score} 
      ] 
      ]
    \addplot[ybar, fill=blue!50!white, draw=blue!50!black] coordinates {(B,0.94) (Ph,0.54) (Sc,0.74)};
    \addplot[ybar, fill=red!50!white, draw=red!50!black] coordinates {(B,0.89) (Ph,0.00) (Sc,0.00)};
\end{axis}

   \end{tikzpicture}
   }
   \vspace{-3mm}
   \caption{Performance comparison of RF, DT, and KNN under baseline and adversarial conditions using the \underline{FGSM} on \underline{Benign} class. The figure illustrates the Precision, Recall, and F1 Score metrics across three classes: \underline{B}enign, \underline{Ph}ishing, and \underline{Sc}amming.}
   \label{fig:Table-9}
   \vspace{-3mm}
\end{figure*}
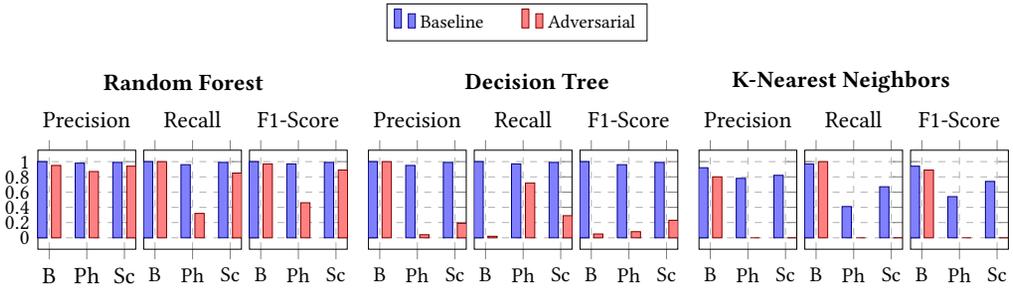

{\noindent \ding{172} \em Benign Class}
The overall accuracy of the RF model dropped from 0.99 to 0.94, with a notable decline in phishing detection. However, it maintained a high accuracy of 0.99 for benign transactions. This suggests that while RF can still recognize benign transactions under adversarial conditions, its ability to detect phishing attempts is significantly weakened, creating a trade-off between precision and security. In contrast, DT’s accuracy fell sharply from 0.99 to 0.09, with benign recall dropping to 0.02, indicating high vulnerability. This sharp decline suggests that DT is particularly sensitive to gradient-based attacks, which can exploit its decision boundaries more easily than those of other models. KNN retained a perfect benign accuracy of 1.00 but saw its overall accuracy decrease from 0.90 to 0.80, showing its difficulty in identifying phishing and scamming transactions. The model behaved differently under adversarial conditions, as shown in Figures \ref{fig:Table-9} and \ref{fig:Model-Accuracy-table9}. This decline highlights KNN’s vulnerability to adversarial attacks, especially when compared to RF and DT. The results remained consistent across multiple metrics, including precision, recall, and F1 scores, with clear performance drops in phishing and scamming detection when exposed to adversarial changes such as FGSM.

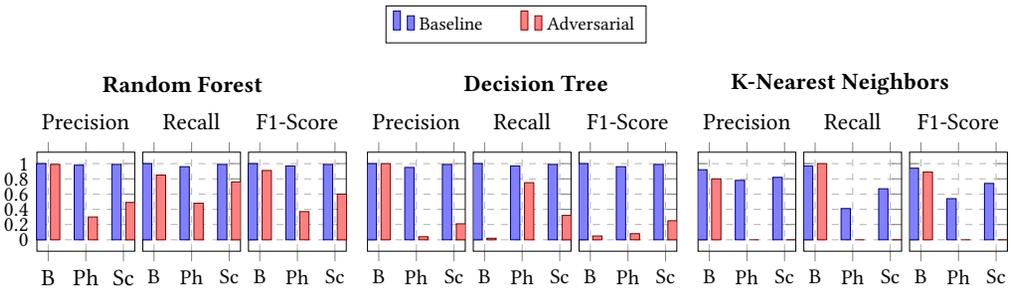
\begin{figure*}[t!]  
     \raggedright
   \vspace{0.2cm}  
    \scalebox{0.88}{
   \begin{tikzpicture}

             \node at (-1.2, 1.5) [xshift=2.4cm]  {\textbf{Random Forest }};

       \begin{axis}[
          at={(-1cm,-1cm)},  
    anchor=south west, 
           ybar,
           ymin = 0, ymax = 1,
           ytick distance = 0.2,
           enlargelimits=0.15,
                     width = 0.22\textwidth,
           height = 0.22\textwidth,
           ylabel style ={font = \fontsize{9}{9}\selectfont},
           yticklabel style={font = \fontsize{9}{9}\selectfont},
            symbolic x coords={B, Ph, Sc},
           xtick=data,
            xtick=data,
           bar width=4pt,
           grid=both,grid style={line width=0.1pt, draw=gray!10},major grid style={line width=.1pt,draw=black!30, dashed},
           title={Precision},
           legend style={at={(4.9, 2.10)}, anchor=south, legend columns=2, font=\fontsize{8}{10}\selectfont, /tikz/every even column/.append style={column sep=0.5cm}}, 
           legend cell align={left}
         ] 
           \addplot[ybar, fill=blue!50!white, draw=blue!50!black] coordinates {(B,1.00) (Ph,0.98) (Sc,0.99)};
           \addlegendentry{Baseline};
           \addplot[ybar, fill=red!50!white, draw=red!50!black] coordinates {(B,0.99) (Ph,0.30) (Sc,0.49)};
           \addlegendentry{Adversarial};
       \end{axis}

\begin{axis}[
    at={(0.6cm,-1cm)}, anchor=south west, 
    ybar,
    ymin = 0, ymax = 1,
    ytick distance = 0.2,
    enlargelimits=0.15,
 width = 0.22\textwidth,
    height = 0.22\textwidth,
    ylabel style ={font = \fontsize{9}{9}\selectfont},
    yticklabel style={font = \fontsize{9}{9}\selectfont},
    yticklabels={}, 
    symbolic x coords={B, Ph, Sc},
    xtick=data,
    bar width=4pt,
    xticklabel style={font = \fontsize{9}{9}\selectfont},
    grid=both,grid style={line width=0.1pt, draw=gray!10},major grid style={line width=0.1pt,draw=black!30, dashed},
    title={Recall}
     ]             
           \addplot[ybar, fill=blue!50!white, draw=blue!50!black] coordinates {(B,1.00) (Ph,0.96) (Sc,0.99)};
           \addplot[ybar, fill=red!50!white, draw=red!50!black] coordinates {(B,0.85) (Ph,0.48) (Sc,0.76)};
       \end{axis}

\begin{axis}[
    at={(2.2cm,-1cm)}, anchor=south west,
    ybar,
    ymin = 0, ymax = 1,
    ytick distance = 0.2,
    enlargelimits=0.15,
 width = 0.22\textwidth,
    height = 0.22\textwidth,
    ylabel style ={font = \fontsize{9}{9}\selectfont},
    yticklabel style={font = \fontsize{9}{9}\selectfont},
    yticklabels={}, 
    symbolic x coords={B, Ph, Sc},
    xtick=data,
    bar width=4pt,
    xticklabel style={font = \fontsize{9}{9}\selectfont},
    grid=both,grid style={line width=0.1pt, draw=gray!10},major grid style={line width=0.1pt,draw=black!30, dashed},
    title={F1-Score}
]             
           \addplot[ybar, fill=blue!50!white, draw=blue!50!black] coordinates {(B,1.00) (Ph,0.97) (Sc,0.99)};
           \addplot[ybar, fill=red!50!white, draw=red!50!black] coordinates {(B,0.91) (Ph,0.37) (Sc,0.60)};
       \end{axis}



       
       
       \node at (-2, 1.5) [xshift=8.55cm]  {\textbf{Decision Tree }};

         \begin{axis}[
    at={(4cm,-1cm)}, anchor=south west,
           ybar,
           ymin = 0, ymax = 1,
           ytick distance = 0.2,
           enlargelimits=0.15,
 width = 0.22\textwidth,
    height = 0.22\textwidth,
    ylabel style ={font = \fontsize{9}{9}\selectfont},
    yticklabel style={font = \fontsize{9}{9}\selectfont},
    yticklabels={}, 
    symbolic x coords={B, Ph, Sc},
    xtick=data,
    bar width=4pt,
    xticklabel style={font = \fontsize{9}{9}\selectfont},
    grid=both,grid style={line width=0.1pt, draw=gray!10},major grid style={line width=0.1pt,draw=black!30, dashed},
    title={Precision}
                 ]       
         ] 
            
           \addplot[ybar, fill=blue!50!white, draw=blue!50!black] coordinates {(B,1.00) (Ph,0.95) (Sc,0.99)};
           \addplot[ybar, fill=red!50!white, draw=red!50!black] coordinates {(B,1.00) (Ph,0.04) (Sc,0.21)};
       \end{axis}

 \begin{axis}[
    at={(5.6cm,-1cm)}, anchor=south west,
    ybar,
    ymin = 0, ymax = 1,
    ytick distance = 0.2,
    enlargelimits=0.15,
 width = 0.22\textwidth,
    height = 0.22\textwidth,
    ylabel style ={font = \fontsize{9}{9}\selectfont},
    yticklabel style={font = \fontsize{9}{9}\selectfont},
    yticklabels={}, 
    symbolic x coords={B, Ph, Sc},
    xtick=data,
    bar width=4pt,
    xticklabel style={font = \fontsize{9}{9}\selectfont},
    grid=both,grid style={line width=0.1pt, draw=gray!10},major grid style={line width=0.1pt,draw=black!30, dashed},
    title={Recall}
      ]       
         ]         
           \addplot[ybar, fill=blue!50!white, draw=blue!50!black] coordinates {(B,1.00) (Ph,0.97) (Sc,0.99)};
           \addplot[ybar, fill=red!50!white, draw=red!50!black] coordinates {(B,0.02) (Ph,0.75) (Sc,0.32)};
       \end{axis}

\begin{axis}[
    at={(7.2cm,-1cm)}, anchor=south west,
    ybar,
    ymin = 0, ymax = 1,
    ytick distance = 0.2,
    enlargelimits=0.15,
 width = 0.22\textwidth,
    height = 0.22\textwidth,
    ylabel style ={font = \fontsize{9}{9}\selectfont},
    yticklabel style={font = \fontsize{9}{9}\selectfont},
    yticklabels={}, 
    symbolic x coords={B, Ph, Sc},
    xtick=data,
    bar width=4pt,
    xticklabel style={font = \fontsize{9}{9}\selectfont},
    grid=both,grid style={line width=0.1pt, draw=gray!10},major grid style={line width=0.1pt,draw=black!30, dashed},
    title={F1-Score} 
      ]       
         ]      
           \addplot[ybar, fill=blue!50!white, draw=blue!50!black] coordinates {(B,1.00) (Ph,0.96) (Sc,0.99)};
           \addplot[ybar, fill=red!50!white, draw=red!50!black] coordinates {(B,0.05) (Ph,0.08) (Sc,0.25)};
       \end{axis}


       \node at (-2, 1.5) [xshift=13.15cm]  {\textbf{K-Nearest Neighbors}};
\begin{axis}[
    at={(9.0cm,-1cm)}, anchor=south west,
    ybar,
    ymin = 0, ymax = 1,
    ytick distance = 0.2,
    enlargelimits=0.15,
 width = 0.22\textwidth,
    height = 0.22\textwidth,
    ylabel style ={font = \fontsize{9}{9}\selectfont},
    yticklabel style={font = \fontsize{9}{9}\selectfont},
    yticklabels={}, 
    symbolic x coords={B, Ph, Sc},
    xtick=data,
    bar width=4pt,
    xticklabel style={font = \fontsize{9}{9}\selectfont},
    grid=both,grid style={line width=0.1pt, draw=gray!10},major grid style={line width=0.1pt,draw=black!30, dashed},
    title={Precision} 
      ]       
         ]       
    \addplot[ybar, fill=blue!50!white, draw=blue!50!black] coordinates {(B,0.92) (Ph,0.78) (Sc,0.82)};
    \addplot[ybar, fill=red!50!white, draw=red!50!black] coordinates {(B,0.80) (Ph,0.00) (Sc,0.00)};
\end{axis}

\begin{axis}[
    at={(10.6cm,-1cm)}, anchor=south west,
    ybar,
    ymin = 0, ymax = 1,
    ytick distance = 0.2,
    enlargelimits=0.15,
 width = 0.22\textwidth,
    height = 0.22\textwidth,
    ylabel style ={font = \fontsize{9}{9}\selectfont},
    yticklabel style={font = \fontsize{9}{9}\selectfont},
    yticklabels={}, 
    symbolic x coords={B, Ph, Sc},
    xtick=data,
    bar width=4pt,
    xticklabel style={font = \fontsize{9}{9}\selectfont},
    grid=both,grid style={line width=0.1pt, draw=gray!10},major grid style={line width=0.1pt,draw=black!30, dashed},
    title={Recall} 
      ] 
]         
    \addplot[ybar, fill=blue!50!white, draw=blue!50!black] coordinates {(B,0.97) (Ph,0.41) (Sc,0.67)};
    \addplot[ybar, fill=red!50!white, draw=red!50!black] coordinates {(B,1.00) (Ph,0.00) (Sc,0.00)};
\end{axis}

\begin{axis}[
     at={(12.2cm,-1cm)}, anchor=south west,
    ybar,
    ymin = 0, ymax = 1,
    ytick distance = 0.2,
    enlargelimits=0.15,
 width = 0.22\textwidth,
    height = 0.22\textwidth,
    ylabel style ={font = \fontsize{9}{9}\selectfont},
    yticklabel style={font = \fontsize{9}{9}\selectfont},
    yticklabels={}, 
    symbolic x coords={B, Ph, Sc},
    xtick=data,
    bar width=4pt,
    xticklabel style={font = \fontsize{9}{9}\selectfont},
    grid=both,grid style={line width=0.1pt, draw=gray!10},major grid style={line width=0.1pt,draw=black!30, dashed},
    title={F1-Score} 
      ] 
      ]
    \addplot[ybar, fill=blue!50!white, draw=blue!50!black] coordinates {(B,0.94) (Ph,0.54) (Sc,0.74)};
    \addplot[ybar, fill=red!50!white, draw=red!50!black] coordinates {(B,0.89) (Ph,0.00) (Sc,0.00)};
\end{axis}
   \end{tikzpicture}

   }
    \vspace{-3mm}
    
   \caption{Evaluation of RF, DT, and KNN performance with \underline{FGSM} on the \underline{Phishing} class. Metrics cover precision, recall, and F1 scores for the \underline{B}enign, \underline{Ph}ishing, and \underline{Sc}amming classes.}
   \label{fig:Table-10}
   
\end{figure*}

{\noindent \ding{173} \em Phishing Class}
The RF model’s phishing detection accuracy dropped from 0.96 to 0.47, with a significant decline in the F1 score. While RF still identified some phishing attempts under attack, its performance was far from reliable, highlighting the need for stronger defense mechanisms. DT’s overall accuracy fell to 0.10, with phishing recall decreasing to 0.75, showing high sensitivity to adversarial attacks. This suggests a structural weakness in DT, making it easier for attacks to manipulate their predictions, particularly for phishing detection. KNN failed entirely to detect phishing transactions under adversarial conditions, with all related metrics dropping to zero. This result shows that KNN becomes ineffective against phishing threats when adversarial modifications are introduced. Figures \ref{fig:Table-10} and \ref{fig:Model-Accuracy-table10} provide a detailed view of how FGSM impacts phishing detection across different models.


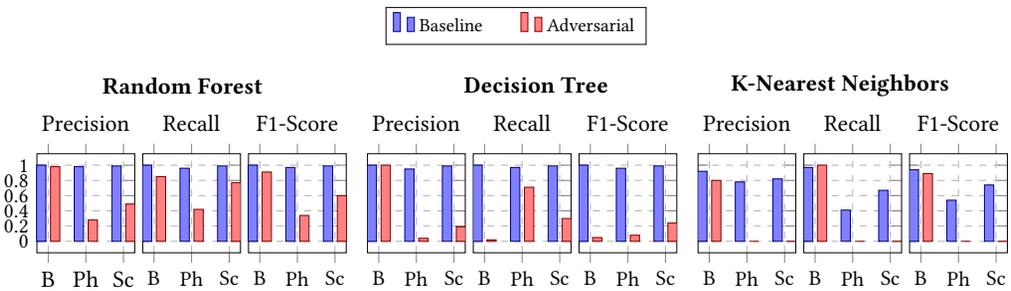
\begin{figure*}[t!]  
     \raggedright
   \vspace{0.2cm}  
    \scalebox{0.88}{
   \begin{tikzpicture}



             \node at (-1.2, 1.5) [xshift=2.4cm]  {\textbf{Random Forest }};

       \begin{axis}[
          at={(-1cm,-1cm)},  
    anchor=south west, 
           ybar,
           ymin = 0, ymax = 1,
           ytick distance = 0.2,
           enlargelimits=0.15,
                     width = 0.22\textwidth,
           height = 0.22\textwidth,
           ylabel style ={font = \fontsize{9}{9}\selectfont},
           yticklabel style={font = \fontsize{9}{9}\selectfont},
            symbolic x coords={B, Ph, Sc},
           xtick=data,
            xtick=data,
           bar width=4pt,
           grid=both,grid style={line width=0.1pt, draw=gray!10},major grid style={line width=.1pt,draw=black!30, dashed},
           title={Precision},
           legend style={at={(4.9, 2.10)}, anchor=south, legend columns=2, font=\fontsize{8}{10}\selectfont, /tikz/every even column/.append style={column sep=0.5cm}}, 
           legend cell align={left}
         ] 
           \addplot[ybar, fill=blue!50!white, draw=blue!50!black] coordinates {(B,1.00) (Ph,0.98) (Sc,0.99)};
           \addlegendentry{Baseline};
           \addplot[ybar, fill=red!50!white, draw=red!50!black] coordinates {(B,0.98) (Ph,0.28) (Sc,0.49)};
           \addlegendentry{Adversarial};
       \end{axis}

\begin{axis}[
    at={(0.6cm,-1cm)}, anchor=south west, 
    ybar,
    ymin = 0, ymax = 1,
    ytick distance = 0.2,
    enlargelimits=0.15,
 width = 0.22\textwidth,
    height = 0.22\textwidth,
    ylabel style ={font = \fontsize{9}{9}\selectfont},
    yticklabel style={font = \fontsize{9}{9}\selectfont},
    yticklabels={}, 
    symbolic x coords={B, Ph, Sc},
    xtick=data,
    bar width=4pt,
    xticklabel style={font = \fontsize{9}{9}\selectfont},
    grid=both,grid style={line width=0.1pt, draw=gray!10},major grid style={line width=0.1pt,draw=black!30, dashed},
    title={Recall}
     ]      
           \addplot[ybar, fill=blue!50!white, draw=blue!50!black] coordinates {(B,1.00) (Ph,0.96) (Sc,0.99)};
           \addplot[ybar, fill=red!50!white, draw=red!50!black] coordinates {(B,0.85) (Ph,0.42) (Sc,0.77)};
       \end{axis}

\begin{axis}[
    at={(2.2cm,-1cm)}, anchor=south west,
    ybar,
    ymin = 0, ymax = 1,
    ytick distance = 0.2,
    enlargelimits=0.15,
 width = 0.22\textwidth,
    height = 0.22\textwidth,
    ylabel style ={font = \fontsize{9}{9}\selectfont},
    yticklabel style={font = \fontsize{9}{9}\selectfont},
    yticklabels={}, 
    symbolic x coords={B, Ph, Sc},
    xtick=data,
    bar width=4pt,
    xticklabel style={font = \fontsize{9}{9}\selectfont},
    grid=both,grid style={line width=0.1pt, draw=gray!10},major grid style={line width=0.1pt,draw=black!30, dashed},
    title={F1-Score}
]         
           \addplot[ybar, fill=blue!50!white, draw=blue!50!black] coordinates {(B,1.00) (Ph,0.97) (Sc,0.99)};
           \addplot[ybar, fill=red!50!white, draw=red!50!black] coordinates {(B,0.91) (Ph,0.34) (Sc,0.60)};
       \end{axis}



       
       \node at (-2, 1.5) [xshift=8.55cm]  {\textbf{Decision Tree }};

         \begin{axis}[
    at={(4cm,-1cm)}, anchor=south west,
           ybar,
           ymin = 0, ymax = 1,
           ytick distance = 0.2,
           enlargelimits=0.15,
 width = 0.22\textwidth,
    height = 0.22\textwidth,
    ylabel style ={font = \fontsize{9}{9}\selectfont},
    yticklabel style={font = \fontsize{9}{9}\selectfont},
    yticklabels={}, 
    symbolic x coords={B, Ph, Sc},
    xtick=data,
    bar width=4pt,
    xticklabel style={font = \fontsize{9}{9}\selectfont},
    grid=both,grid style={line width=0.1pt, draw=gray!10},major grid style={line width=0.1pt,draw=black!30, dashed},
    title={Precision}
                 ]       
         ] 
            
           \addplot[ybar, fill=blue!50!white, draw=blue!50!black] coordinates {(B,1.00) (Ph,0.95) (Sc,0.99)};
           \addplot[ybar, fill=red!50!white, draw=red!50!black] coordinates {(B,1.00) (Ph,0.04) (Sc,0.19)};
       \end{axis}
       
\begin{axis}[
    at={(5.6cm,-1cm)}, anchor=south west,
    ybar,
    ymin = 0, ymax = 1,
    ytick distance = 0.2,
    enlargelimits=0.15,
 width = 0.22\textwidth,
    height = 0.22\textwidth,
    ylabel style ={font = \fontsize{9}{9}\selectfont},
    yticklabel style={font = \fontsize{9}{9}\selectfont},
    yticklabels={}, 
    symbolic x coords={B, Ph, Sc},
    xtick=data,
    bar width=4pt,
    xticklabel style={font = \fontsize{9}{9}\selectfont},
    grid=both,grid style={line width=0.1pt, draw=gray!10},major grid style={line width=0.1pt,draw=black!30, dashed},
    title={Recall}
      ]       
         ]        
           \addplot[ybar, fill=blue!50!white, draw=blue!50!black] coordinates {(B,1.00) (Ph,0.97) (Sc,0.99)};
           \addplot[ybar, fill=red!50!white, draw=red!50!black] coordinates {(B,0.02) (Ph,0.71) (Sc,0.30)};
       \end{axis}

\begin{axis}[
    at={(7.2cm,-1cm)}, anchor=south west,
    ybar,
    ymin = 0, ymax = 1,
    ytick distance = 0.2,
    enlargelimits=0.15,
 width = 0.22\textwidth,
    height = 0.22\textwidth,
    ylabel style ={font = \fontsize{9}{9}\selectfont},
    yticklabel style={font = \fontsize{9}{9}\selectfont},
    yticklabels={}, 
    symbolic x coords={B, Ph, Sc},
    xtick=data,
    bar width=4pt,
    xticklabel style={font = \fontsize{9}{9}\selectfont},
    grid=both,grid style={line width=0.1pt, draw=gray!10},major grid style={line width=0.1pt,draw=black!30, dashed},
    title={F1-Score} 
      ]       
         ]       
           \addplot[ybar, fill=blue!50!white, draw=blue!50!black] coordinates {(B,1.00) (Ph,0.96) (Sc,0.99)};
           \addplot[ybar, fill=red!50!white, draw=red!50!black] coordinates {(B,0.05) (Ph,0.08) (Sc,0.24)};
       \end{axis}


       \node at (-2, 1.5) [xshift=13.15cm]  {\textbf{K-Nearest Neighbors}};
\begin{axis}[
    at={(9.0cm,-1cm)}, anchor=south west,
    ybar,
    ymin = 0, ymax = 1,
    ytick distance = 0.2,
    enlargelimits=0.15,
 width = 0.22\textwidth,
    height = 0.22\textwidth,
    ylabel style ={font = \fontsize{9}{9}\selectfont},
    yticklabel style={font = \fontsize{9}{9}\selectfont},
    yticklabels={}, 
    symbolic x coords={B, Ph, Sc},
    xtick=data,
    bar width=4pt,
    xticklabel style={font = \fontsize{9}{9}\selectfont},
    grid=both,grid style={line width=0.1pt, draw=gray!10},major grid style={line width=0.1pt,draw=black!30, dashed},
    title={Precision} 
      ]       
         ]   
    \addplot[ybar, fill=blue!50!white, draw=blue!50!black] coordinates {(B,0.92) (Ph,0.78) (Sc,0.82)};
    \addplot[ybar, fill=red!50!white, draw=red!50!black] coordinates {(B,0.80) (Ph,0.00) (Sc,0.00)};
\end{axis}

\begin{axis}[
    at={(10.6cm,-1cm)}, anchor=south west,
    ybar,
    ymin = 0, ymax = 1,
    ytick distance = 0.2,
    enlargelimits=0.15,
 width = 0.22\textwidth,
    height = 0.22\textwidth,
    ylabel style ={font = \fontsize{9}{9}\selectfont},
    yticklabel style={font = \fontsize{9}{9}\selectfont},
    yticklabels={}, 
    symbolic x coords={B, Ph, Sc},
    xtick=data,
    bar width=4pt,
    xticklabel style={font = \fontsize{9}{9}\selectfont},
    grid=both,grid style={line width=0.1pt, draw=gray!10},major grid style={line width=0.1pt,draw=black!30, dashed},
    title={Recall} 
      ] 
]      
    \addplot[ybar, fill=blue!50!white, draw=blue!50!black] coordinates {(B,0.97) (Ph,0.41) (Sc,0.67)};
    \addplot[ybar, fill=red!50!white, draw=red!50!black] coordinates {(B,1.00) (Ph,0.00) (Sc,0.00)};
\end{axis}

\begin{axis}[
     at={(12.2cm,-1cm)}, anchor=south west,
    ybar,
    ymin = 0, ymax = 1,
    ytick distance = 0.2,
    enlargelimits=0.15,
 width = 0.22\textwidth,
    height = 0.22\textwidth,
    ylabel style ={font = \fontsize{9}{9}\selectfont},
    yticklabel style={font = \fontsize{9}{9}\selectfont},
    yticklabels={}, 
    symbolic x coords={B, Ph, Sc},
    xtick=data,
    bar width=4pt,
    xticklabel style={font = \fontsize{9}{9}\selectfont},
    grid=both,grid style={line width=0.1pt, draw=gray!10},major grid style={line width=0.1pt,draw=black!30, dashed},
    title={F1-Score} 
      ] 
      ]   
    \addplot[ybar, fill=blue!50!white, draw=blue!50!black] coordinates {(B,0.94) (Ph,0.54) (Sc,0.74)};
    \addplot[ybar, fill=red!50!white, draw=red!50!black] coordinates {(B,0.89) (Ph,0.00) (Sc,0.00)};
\end{axis}
   \end{tikzpicture}
   }
   \vspace{-3mm}
   
   \caption{Assessment of RF, DT, and KNN performance with \underline{FGSM} on the \underline{Scamming} class. Metrics include overall and scamming class accuracy, as well as precision, recall, and F1 scores for \underline{B}enign, \underline{Ph}ishing, and \underline{Sc}amming across baseline and adversarial conditions.}
   \label{fig:Table-11}
\end{figure*}

{\noindent \ding{174} \em Scamming Class}
The overall accuracy of the RF model dropped from 0.99 to 0.81, with scamming accuracy decreasing from 0.99 to 0.76. While RF still outperforms other models in this scenario, the significant drop indicates a need for improved resistance to adversarial attacks in scam detection. The DT’s overall accuracy fell to 0.09, with scamming recall drastically reducing to 0.30. The DT model's performance suggests it is highly vulnerable to adversarial manipulations in scam-related features, making it less reliable for detecting such fraudulent activities. The KNN model’s scamming detection metrics also dropped to zero. Figures \ref{fig:Table-11} and \ref{fig:Model-Accuracy-table11} depicts a comparison between the baseline and adversarial performance using FGSM on the scamming class across different models.


\takeaway{Gradient-based adversarial attacks, such as FGSM, severely compromise the effectiveness of machine learning models in detecting phishing and scamming activities, particularly for Decision Tree and KNN models, which become nearly ineffective under such conditions.}

\subsection{Results of Untargeted Attacks}

We employed untargeted adversarial attacks to evaluate the models' performance in adversarial scenarios. This method involved modifying different features to examine which models could sustain their detection capabilities when exposed to manipulated data. Rather than concentrating on specific targeted attacks, the primary objective was to assess the overall robustness of each model against a diverse range of perturbations. 

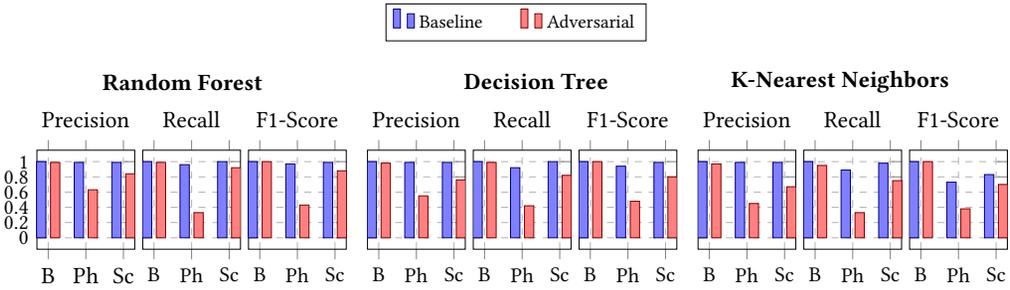
\begin{figure*}[t!]  
     \raggedright
   \vspace{0.2cm}  
    \scalebox{0.88}{
   \begin{tikzpicture}



             \node at (-1.2, 1.5) [xshift=2.4cm]  {\textbf{Random Forest }};

       \begin{axis}[
          at={(-1cm,-1cm)},  
    anchor=south west, 
           ybar,
           ymin = 0, ymax = 1,
           ytick distance = 0.2,
           enlargelimits=0.15,
                     width = 0.22\textwidth,
           height = 0.22\textwidth,
           ylabel style ={font = \fontsize{9}{9}\selectfont},
           yticklabel style={font = \fontsize{9}{9}\selectfont},
            symbolic x coords={B, Ph, Sc},
           xtick=data,
            xtick=data,
           bar width=4pt,
           grid=both,grid style={line width=0.1pt, draw=gray!10},major grid style={line width=.1pt,draw=black!30, dashed},
           title={Precision},
           legend style={at={(4.9, 2.10)}, anchor=south, legend columns=2, font=\fontsize{8}{10}\selectfont, /tikz/every even column/.append style={column sep=0.5cm}}, 
           legend cell align={left}
         ]      
           \addplot[ybar, fill=blue!50!white, draw=blue!50!black] coordinates {(B,1.00) (Ph,0.99) (Sc,0.99)};
           \addlegendentry{Baseline};
           \addplot[ybar, fill=red!50!white, draw=red!50!black] coordinates {(B,0.99) (Ph,0.63) (Sc,0.84)};
           \addlegendentry{Adversarial};
       \end{axis}

\begin{axis}[
    at={(0.6cm,-1cm)}, anchor=south west, 
    ybar,
    ymin = 0, ymax = 1,
    ytick distance = 0.2,
    enlargelimits=0.15,
 width = 0.22\textwidth,
    height = 0.22\textwidth,
    ylabel style ={font = \fontsize{9}{9}\selectfont},
    yticklabel style={font = \fontsize{9}{9}\selectfont},
    yticklabels={}, 
    symbolic x coords={B, Ph, Sc},
    xtick=data,
    bar width=4pt,
    xticklabel style={font = \fontsize{9}{9}\selectfont},
    grid=both,grid style={line width=0.1pt, draw=gray!10},major grid style={line width=0.1pt,draw=black!30, dashed},
    title={Recall}
     ]        
    \addplot[ybar, fill=blue!50!white, draw=blue!50!black] coordinates {(B,1.00) (Ph,0.96) (Sc,1.00)};
    \addplot[ybar, fill=red!50!white, draw=red!50!black] coordinates {(B,0.99) (Ph,0.33) (Sc,0.92)};
\end{axis}

\begin{axis}[
    at={(2.2cm,-1cm)}, anchor=south west,
    ybar,
    ymin = 0, ymax = 1,
    ytick distance = 0.2,
    enlargelimits=0.15,
 width = 0.22\textwidth,
    height = 0.22\textwidth,
    ylabel style ={font = \fontsize{9}{9}\selectfont},
    yticklabel style={font = \fontsize{9}{9}\selectfont},
    yticklabels={}, 
    symbolic x coords={B, Ph, Sc},
    xtick=data,
    bar width=4pt,
    xticklabel style={font = \fontsize{9}{9}\selectfont},
    grid=both,grid style={line width=0.1pt, draw=gray!10},major grid style={line width=0.1pt,draw=black!30, dashed},
    title={F1-Score}
]         
    \addplot[ybar, fill=blue!50!white, draw=blue!50!black] coordinates {(B,1.00) (Ph,0.97) (Sc,0.99)};
    \addplot[ybar, fill=red!50!white, draw=red!50!black] coordinates {(B,1.00) (Ph,0.43) (Sc,0.88)};
\end{axis}



       
       
       \node at (-2, 1.5) [xshift=8.55cm]  {\textbf{Decision Tree }};

         \begin{axis}[
    at={(4cm,-1cm)}, anchor=south west,
           ybar,
           ymin = 0, ymax = 1,
           ytick distance = 0.2,
           enlargelimits=0.15,
 width = 0.22\textwidth,
    height = 0.22\textwidth,
    ylabel style ={font = \fontsize{9}{9}\selectfont},
    yticklabel style={font = \fontsize{9}{9}\selectfont},
    yticklabels={}, 
    symbolic x coords={B, Ph, Sc},
    xtick=data,
    bar width=4pt,
    xticklabel style={font = \fontsize{9}{9}\selectfont},
    grid=both,grid style={line width=0.1pt, draw=gray!10},major grid style={line width=0.1pt,draw=black!30, dashed},
    title={Precision}
                 ]       
         ]     
           \addplot[ybar, fill=blue!50!white, draw=blue!50!black] coordinates {(B,1.00) (Ph,0.99) (Sc,0.99)};
           \addplot[ybar, fill=red!50!white, draw=red!50!black] coordinates {(B,0.98) (Ph,0.55) (Sc,0.76)};
       \end{axis}

\begin{axis}[
    at={(5.6cm,-1cm)}, anchor=south west,
    ybar,
    ymin = 0, ymax = 1,
    ytick distance = 0.2,
    enlargelimits=0.15,
 width = 0.22\textwidth,
    height = 0.22\textwidth,
    ylabel style ={font = \fontsize{9}{9}\selectfont},
    yticklabel style={font = \fontsize{9}{9}\selectfont},
    yticklabels={}, 
    symbolic x coords={B, Ph, Sc},
    xtick=data,
    bar width=4pt,
    xticklabel style={font = \fontsize{9}{9}\selectfont},
    grid=both,grid style={line width=0.1pt, draw=gray!10},major grid style={line width=0.1pt,draw=black!30, dashed},
    title={Recall}
      ]       
         ]        
    \addplot[ybar, fill=blue!50!white, draw=blue!50!black] coordinates {(B,1.00) (Ph,0.92) (Sc,1.00)};
    \addplot[ybar, fill=red!50!white, draw=red!50!black] coordinates {(B,0.99) (Ph,0.42) (Sc,0.82)};
\end{axis}

\begin{axis}[
    at={(7.2cm,-1cm)}, anchor=south west,
    ybar,
    ymin = 0, ymax = 1,
    ytick distance = 0.2,
    enlargelimits=0.15,
 width = 0.22\textwidth,
    height = 0.22\textwidth,
    ylabel style ={font = \fontsize{9}{9}\selectfont},
    yticklabel style={font = \fontsize{9}{9}\selectfont},
    yticklabels={}, 
    symbolic x coords={B, Ph, Sc},
    xtick=data,
    bar width=4pt,
    xticklabel style={font = \fontsize{9}{9}\selectfont},
    grid=both,grid style={line width=0.1pt, draw=gray!10},major grid style={line width=0.1pt,draw=black!30, dashed},
    title={F1-Score} 
      ]       
         ]    
    \addplot[ybar, fill=blue!50!white, draw=blue!50!black] coordinates {(B,1.00) (Ph,0.94) (Sc,0.99)};
    \addplot[ybar, fill=red!50!white, draw=red!50!black] coordinates {(B,1.00) (Ph,0.48) (Sc,0.80)};
\end{axis}


       \node at (-2, 1.5) [xshift=13.15cm]  {\textbf{K-Nearest Neighbors}};
\begin{axis}[
    at={(9.0cm,-1cm)}, anchor=south west,
    ybar,
    ymin = 0, ymax = 1,
    ytick distance = 0.2,
    enlargelimits=0.15,
 width = 0.22\textwidth,
    height = 0.22\textwidth,
    ylabel style ={font = \fontsize{9}{9}\selectfont},
    yticklabel style={font = \fontsize{9}{9}\selectfont},
    yticklabels={}, 
    symbolic x coords={B, Ph, Sc},
    xtick=data,
    bar width=4pt,
    xticklabel style={font = \fontsize{9}{9}\selectfont},
    grid=both,grid style={line width=0.1pt, draw=gray!10},major grid style={line width=0.1pt,draw=black!30, dashed},
    title={Precision} 
      ]       
         ]       
    \addplot[ybar, fill=blue!50!white, draw=blue!50!black] coordinates {(B,1.00) (Ph,0.99) (Sc,0.99)};
    \addplot[ybar, fill=red!50!white, draw=red!50!black] coordinates {(B,0.97) (Ph,0.45) (Sc,0.67)};
\end{axis}

\begin{axis}[
    at={(10.6cm,-1cm)}, anchor=south west,
    ybar,
    ymin = 0, ymax = 1,
    ytick distance = 0.2,
    enlargelimits=0.15,
 width = 0.22\textwidth,
    height = 0.22\textwidth,
    ylabel style ={font = \fontsize{9}{9}\selectfont},
    yticklabel style={font = \fontsize{9}{9}\selectfont},
    yticklabels={}, 
    symbolic x coords={B, Ph, Sc},
    xtick=data,
    bar width=4pt,
    xticklabel style={font = \fontsize{9}{9}\selectfont},
    grid=both,grid style={line width=0.1pt, draw=gray!10},major grid style={line width=0.1pt,draw=black!30, dashed},
    title={Recall} 
      ] 
]        
    \addplot[ybar, fill=blue!50!white, draw=blue!50!black] coordinates {(B,1.00) (Ph,0.89) (Sc,0.98)};
    \addplot[ybar, fill=red!50!white, draw=red!50!black] coordinates {(B,0.95) (Ph,0.33) (Sc,0.75)};
\end{axis}

\begin{axis}[
     at={(12.2cm,-1cm)}, anchor=south west,
    ybar,
    ymin = 0, ymax = 1,
    ytick distance = 0.2,
    enlargelimits=0.15,
 width = 0.22\textwidth,
    height = 0.22\textwidth,
    ylabel style ={font = \fontsize{9}{9}\selectfont},
    yticklabel style={font = \fontsize{9}{9}\selectfont},
    yticklabels={}, 
    symbolic x coords={B, Ph, Sc},
    xtick=data,
    bar width=4pt,
    xticklabel style={font = \fontsize{9}{9}\selectfont},
    grid=both,grid style={line width=0.1pt, draw=gray!10},major grid style={line width=0.1pt,draw=black!30, dashed},
    title={F1-Score} 
      ] 
      ]       
    \addplot[ybar, fill=blue!50!white, draw=blue!50!black] coordinates {(B,1.00) (Ph,0.73) (Sc,0.83)};
    \addplot[ybar, fill=red!50!white, draw=red!50!black] coordinates {(B,1.00) (Ph,0.38) (Sc,0.70)};
\end{axis}
   \end{tikzpicture}
   }
\vspace{-3mm}   
   \caption{Performance comparison of RF, DT, and KNN models under baseline and untargeted adversarial attacks. The plots illustrate Precision, Recall, and F1-Score across \underline{B}enign, \underline{Ph}ishing, and \underline{Sc}amming.}

   \label{fig:Model-Metrics-table12}
\end{figure*}

{\noindent \ding{172} \em All Features}
The RF model's accuracy decreased to 0.95, DT’s to 0.91, with phishing detection severely impaired, and KNN maintained its baseline accuracy. These findings indicate that while RF and DT models exhibit robustness, they remain vulnerable to broad, untargeted adversarial modifications. Notably, the phishing precision for RF drops from 0.99 to 0.63, and for DT, from 0.96 to 0.15. Similarly, phishing recall decreases from 0.96 to 0.33 for RF and from 0.97 to 0.15 for DT. These reductions underscore the models' reduced ability to identify phishing attempts when under attack. The details are illustrated in Figures \ref{fig:Model-Metrics-table12} and \ref{fig:Model-Accuracy-table12}, where performance metrics under these conditions are compared across all models. This significant drop in accuracy highlights the susceptibility of these models to broad adversarial manipulations.


{\noindent \ding{173} \em Address Features}
AEs focusing on the {\em from\_address} and {\em to\_address} features resulted in a decline in overall accuracy to 0.80 for all models. None of the models detected phishing or scamming, indicating a high sensitivity to address manipulations. This result highlights the critical importance of accurately processing and interpreting address features, as adversaries can easily exploit these weaknesses to evade detection. The complete drop in phishing and scamming detection metrics to zero across all models is alarming, considering the baseline phishing precision and recall were 0.95 and 0.74 for RF, 0.94 and 0.74 for DT, and 0.88 and 0.76 for KNN, respectively. These declines underscore the vulnerability of these models when address features are manipulated. This vulnerability is further illustrated in Figures \ref{fig:Table-13} and \ref{fig:Model-Accuracy-table13}, demonstrating the impact of unseen address manipulations on model performance.

These findings highlight the importance of accurate processing, as attackers can easily exploit any manipulation to bypass detection. The fact that phishing and scamming detection metrics dropped to zero across all models is alarming. Before manipulation, the baseline phishing precision and recall were 0.95 and 0.74 for RF, 0.94 and 0.74 for DT, and 0.88 and 0.76 for KNN. The drop in performance underscores how vulnerable these models become when key features are altered. The results in \autoref{table:unseen_address_D2},  show the impact of unseen address manipulations on model performance.


\begin{figure*}[t!]  
     \raggedright
   \vspace{0.2cm}  
    \scalebox{0.88}{
   \begin{tikzpicture}



             \node at (-1.2, 1.5) [xshift=2.4cm]  {\textbf{Random Forest }};

       \begin{axis}[
          at={(-1cm,-1cm)},  
    anchor=south west, 
           ybar,
           ymin = 0, ymax = 1,
           ytick distance = 0.2,
           enlargelimits=0.15,
                     width = 0.22\textwidth,
           height = 0.22\textwidth,
           ylabel style ={font = \fontsize{9}{9}\selectfont},
           yticklabel style={font = \fontsize{9}{9}\selectfont},
            symbolic x coords={B, Ph, Sc},
           xtick=data,
            xtick=data,
           bar width=4pt,
           grid=both,grid style={line width=0.1pt, draw=gray!10},major grid style={line width=.1pt,draw=black!30, dashed},
           title={Precision},
           legend style={at={(4.9, 2.10)}, anchor=south, legend columns=2, font=\fontsize{8}{10}\selectfont, /tikz/every even column/.append style={column sep=0.5cm}}, 
           legend cell align={left}
         ]      
         
           \addplot[ybar, fill=blue!50!white, draw=blue!50!black] coordinates {(B,0.98) (Ph,0.95) (Sc,0.99)};
           \addlegendentry{Baseline};
           \addplot[ybar, fill=red!50!white, draw=red!50!black] coordinates {(B,0.80) (Ph,0.00) (Sc,0.00)};
           \addlegendentry{Adversarial};
       \end{axis}

\begin{axis}[
    at={(0.6cm,-1cm)}, anchor=south west, 
    ybar,
    ymin = 0, ymax = 1,
    ytick distance = 0.2,
    enlargelimits=0.15,
 width = 0.22\textwidth,
    height = 0.22\textwidth,
    ylabel style ={font = \fontsize{9}{9}\selectfont},
    yticklabel style={font = \fontsize{9}{9}\selectfont},
    yticklabels={}, 
    symbolic x coords={B, Ph, Sc},
    xtick=data,
    bar width=4pt,
    xticklabel style={font = \fontsize{9}{9}\selectfont},
    grid=both,grid style={line width=0.1pt, draw=gray!10},major grid style={line width=0.1pt,draw=black!30, dashed},
    title={Recall}
     ]   
     \addplot[ybar, fill=blue!50!white, draw=blue!50!black] coordinates {(B,1.00) (Ph,0.74) (Sc,0.93)};
           \addplot[ybar, fill=red!50!white, draw=red!50!black] coordinates {(B,1.00) (Ph,0.00) (Sc,0.00)};
       \end{axis}

\begin{axis}[
    at={(2.2cm,-1cm)}, anchor=south west,
    ybar,
    ymin = 0, ymax = 1,
    ytick distance = 0.2,
    enlargelimits=0.15,
 width = 0.22\textwidth,
    height = 0.22\textwidth,
    ylabel style ={font = \fontsize{9}{9}\selectfont},
    yticklabel style={font = \fontsize{9}{9}\selectfont},
    yticklabels={}, 
    symbolic x coords={B, Ph, Sc},
    xtick=data,
    bar width=4pt,
    xticklabel style={font = \fontsize{9}{9}\selectfont},
    grid=both,grid style={line width=0.1pt, draw=gray!10},major grid style={line width=0.1pt,draw=black!30, dashed},
    title={F1-Score}
]        
           \addplot[ybar, fill=blue!50!white, draw=blue!50!black] coordinates {(B,0.99) (Ph,0.83) (Sc,0.96)};
           \addplot[ybar, fill=red!50!white, draw=red!50!black] coordinates {(B,0.89) (Ph,0.00) (Sc,0.00)};
       \end{axis}



       
       \node at (-2, 1.5) [xshift=8.55cm]  {\textbf{Decision Tree }};

         \begin{axis}[
    at={(4cm,-1cm)}, anchor=south west,
           ybar,
           ymin = 0, ymax = 1,
           ytick distance = 0.2,
           enlargelimits=0.15,
 width = 0.22\textwidth,
    height = 0.22\textwidth,
    ylabel style ={font = \fontsize{9}{9}\selectfont},
    yticklabel style={font = \fontsize{9}{9}\selectfont},
    yticklabels={}, 
    symbolic x coords={B, Ph, Sc},
    xtick=data,
    bar width=4pt,
    xticklabel style={font = \fontsize{9}{9}\selectfont},
    grid=both,grid style={line width=0.1pt, draw=gray!10},major grid style={line width=0.1pt,draw=black!30, dashed},
    title={Precision}
                 ]       
         ]     
             
           \addplot[ybar, fill=blue!50!white, draw=blue!50!black] coordinates {(B,0.97) (Ph,0.94) (Sc,0.99)};
           \addplot[ybar, fill=red!50!white, draw=red!50!black] coordinates {(B,0.80) (Ph,0.00) (Sc,0.00)};
       \end{axis}

\begin{axis}[
    at={(5.6cm,-1cm)}, anchor=south west,
    ybar,
    ymin = 0, ymax = 1,
    ytick distance = 0.2,
    enlargelimits=0.15,
 width = 0.22\textwidth,
    height = 0.22\textwidth,
    ylabel style ={font = \fontsize{9}{9}\selectfont},
    yticklabel style={font = \fontsize{9}{9}\selectfont},
    yticklabels={}, 
    symbolic x coords={B, Ph, Sc},
    xtick=data,
    bar width=4pt,
    xticklabel style={font = \fontsize{9}{9}\selectfont},
    grid=both,grid style={line width=0.1pt, draw=gray!10},major grid style={line width=0.1pt,draw=black!30, dashed},
    title={Recall}
      ]       
         ]           
           \addplot[ybar, fill=blue!50!white, draw=blue!50!black] coordinates {(B,1.00) (Ph,0.74) (Sc,0.93)};
           \addplot[ybar, fill=red!50!white, draw=red!50!black] coordinates {(B,1.00) (Ph,0.00) (Sc,0.00)};
       \end{axis}

\begin{axis}[
    at={(7.2cm,-1cm)}, anchor=south west,
    ybar,
    ymin = 0, ymax = 1,
    ytick distance = 0.2,
    enlargelimits=0.15,
 width = 0.22\textwidth,
    height = 0.22\textwidth,
    ylabel style ={font = \fontsize{9}{9}\selectfont},
    yticklabel style={font = \fontsize{9}{9}\selectfont},
    yticklabels={}, 
    symbolic x coords={B, Ph, Sc},
    xtick=data,
    bar width=4pt,
    xticklabel style={font = \fontsize{9}{9}\selectfont},
    grid=both,grid style={line width=0.1pt, draw=gray!10},major grid style={line width=0.1pt,draw=black!30, dashed},
    title={F1-Score} 
      ]       
         ]         
           \addplot[ybar, fill=blue!50!white, draw=blue!50!black] coordinates {(B,0.99) (Ph,0.83) (Sc,0.96)};
           \addplot[ybar, fill=red!50!white, draw=red!50!black] coordinates {(B,0.89) (Ph,0.00) (Sc,0.00)};
       \end{axis}


       \node at (-2, 1.5) [xshift=13.15cm]  {\textbf{K-Nearest Neighbors}};
\begin{axis}[
    at={(9.0cm,-1cm)}, anchor=south west,
    ybar,
    ymin = 0, ymax = 1,
    ytick distance = 0.2,
    enlargelimits=0.15,
 width = 0.22\textwidth,
    height = 0.22\textwidth,
    ylabel style ={font = \fontsize{9}{9}\selectfont},
    yticklabel style={font = \fontsize{9}{9}\selectfont},
    yticklabels={}, 
    symbolic x coords={B, Ph, Sc},
    xtick=data,
    bar width=4pt,
    xticklabel style={font = \fontsize{9}{9}\selectfont},
    grid=both,grid style={line width=0.1pt, draw=gray!10},major grid style={line width=0.1pt,draw=black!30, dashed},
    title={Precision} 
      ]       
         ]   
    \addplot[ybar, fill=blue!50!white, draw=blue!50!black] coordinates {(B,0.98) (Ph,0.88) (Sc,0.97)};
    \addplot[ybar, fill=red!50!white, draw=red!50!black] coordinates {(B,0.80) (Ph,0.00) (Sc,0.00)};
\end{axis}

\begin{axis}[
    at={(10.6cm,-1cm)}, anchor=south west,
    ybar,
    ymin = 0, ymax = 1,
    ytick distance = 0.2,
    enlargelimits=0.15,
 width = 0.22\textwidth,
    height = 0.22\textwidth,
    ylabel style ={font = \fontsize{9}{9}\selectfont},
    yticklabel style={font = \fontsize{9}{9}\selectfont},
    yticklabels={}, 
    symbolic x coords={B, Ph, Sc},
    xtick=data,
    bar width=4pt,
    xticklabel style={font = \fontsize{9}{9}\selectfont},
    grid=both,grid style={line width=0.1pt, draw=gray!10},major grid style={line width=0.1pt,draw=black!30, dashed},
    title={Recall} 
      ] 
] 
\addplot[ybar, fill=blue!50!white, draw=blue!50!black] coordinates {(B,1.00) (Ph,0.76) (Sc,0.93)};
    \addplot[ybar, fill=red!50!white, draw=red!50!black] coordinates {(B,1.00) (Ph,0.00) (Sc,0.00)};
\end{axis}

\begin{axis}[
     at={(12.2cm,-1cm)}, anchor=south west,
    ybar,
    ymin = 0, ymax = 1,
    ytick distance = 0.2,
    enlargelimits=0.15,
 width = 0.22\textwidth,
    height = 0.22\textwidth,
    ylabel style ={font = \fontsize{9}{9}\selectfont},
    yticklabel style={font = \fontsize{9}{9}\selectfont},
    yticklabels={}, 
    symbolic x coords={B, Ph, Sc},
    xtick=data,
    bar width=4pt,
    xticklabel style={font = \fontsize{9}{9}\selectfont},
    grid=both,grid style={line width=0.1pt, draw=gray!10},major grid style={line width=0.1pt,draw=black!30, dashed},
    title={F1-Score} 
      ] 
      ]          
    \addplot[ybar, fill=blue!50!white, draw=blue!50!black] coordinates {(B,0.99) (Ph,0.81) (Sc,0.95)};
    \addplot[ybar, fill=red!50!white, draw=red!50!black] coordinates {(B,0.89) (Ph,0.00) (Sc,0.00)};
\end{axis}
   \end{tikzpicture}
   }
   \vspace{-3mm}
   \caption{RF, DT, and KNN  performance with address feature manipulation. Metrics: accuracy, precision, recall, and F1 for \underline{B}enign, \underline{Ph}ishing, and \underline{Sc}amming classes.}
   \label{fig:Table-13}
   
\end{figure*}
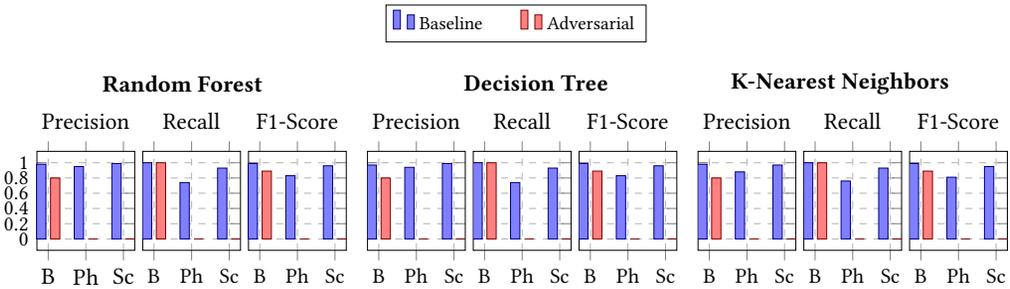

 \begin{figure*}[h]
    \centering
    \vspace{0.2cm}  

    \begin{subfigure}{0.18\textwidth}  
        \centering
        \begin{tikzpicture}
            \begin{axis}[
                ybar,
                ymin=0, ymax=1,
                ytick distance=0.2,
                enlargelimits=0.16,
                width=1.3\textwidth,
                height=1.3\textwidth,  
                ylabel style={font=\fontsize{10}{10}\selectfont},
                yticklabel style={font=\fontsize{6}{6}\selectfont},
                symbolic x coords={RF, DT, kNN},
                xtick=data,
                xticklabel style={font=\fontsize{7}{7}\selectfont},
                bar width=3pt,  
                grid=both,
                grid style={line width=0.1pt, draw=gray!10},
                major grid style={line width=.4pt,draw=black!30, dashed},
                legend style={at={(3.90,1.30)}, anchor=south, legend columns=2, font=\fontsize{8}{8}\selectfont},  
                legend cell align={right}
            ]
                \addplot[ybar, fill=blue!50!white, draw=blue!50!black] coordinates {(RF,0.99) (DT,0.99) (kNN,0.90)};
                \addlegendentry{Baseline};
                \addplot[ybar, fill=red!50!white, draw=red!50!black] coordinates {(RF,0.94) (DT,0.95) (kNN,0.80)};
                \addlegendentry{Adversarial};
            \end{axis}
        \end{tikzpicture}

        \caption{Models accuracy: FGSM effect on benign class. \autoref{fig:Table-9} references the precision, recall, and F1-score metrics.}
        \label{fig:Model-Accuracy-table9}
    \end{subfigure}
    \hfill
    \begin{subfigure}{0.18\textwidth}  
        \centering
        \begin{tikzpicture}
            \begin{axis}[
                ybar,
                ymin=0, ymax=1,
                ytick distance=0.2,
                enlargelimits=0.16,
                width=1.3\textwidth,
                height=1.3\textwidth,  
                ylabel style={font=\fontsize{10}{10}\selectfont},
                yticklabel style={font=\fontsize{6}{6}\selectfont},
                symbolic x coords={RF, DT, KNN},
                xtick=data,
                xticklabel style={font=\fontsize{7}{7}\selectfont},
                bar width=3pt,  
                grid=both,
                grid style={line width=0.1pt, draw=gray!10},
                major grid style={line width=.1pt,draw=black!30, dashed},
            ]
                \addplot[ybar, fill=blue!50!white, draw=blue!50!black] coordinates {(RF,0.99) (DT,0.99) (KNN,0.90)};
                \addplot[ybar, fill=red!50!white, draw=red!50!black] coordinates {(RF,0.81) (DT,0.10) (KNN,0.80)};
            \end{axis}
        \end{tikzpicture}
    
        \caption{Models accuracy: FGSM effect on phishing class.\autoref{fig:Table-10} references the precision, recall, and F1-score metrics.}
        \label{fig:Model-Accuracy-table10}
    \end{subfigure}
    \hfill
    \begin{subfigure}{0.18\textwidth}  
        \centering
        \begin{tikzpicture}
            \begin{axis}[
                ybar,
                ymin=0, ymax=1,
                ytick distance=0.2,
                enlargelimits=0.16,
                 width=1.3\textwidth,
                height=1.3\textwidth, 
                ylabel style={font=\fontsize{10}{10}\selectfont},
                yticklabel style={font=\fontsize{6}{6}\selectfont},
                symbolic x coords={RF, DT, KNN},
                xtick=data,
                xticklabel style={font=\fontsize{7}{7}\selectfont},
                bar width=3pt,  
                grid=both,
                grid style={line width=0.1pt, draw=gray!10},
                major grid style={line width=0.1pt,draw=black!30, dashed},
            ]
                \addplot[ybar, fill=blue!50!white, draw=blue!50!black] coordinates {(RF,0.99) (DT,0.99) (KNN,0.90)};
                \addplot[ybar, fill=red!50!white, draw=red!50!black] coordinates {(RF,0.81) (DT,0.09) (KNN,0.80)};
            \end{axis}
        \end{tikzpicture}

        \caption{Models accuracy: FGSM effect on scamming class. \autoref{fig:Table-11} references the precision, recall, and F1-score metrics.}
        \label{fig:Model-Accuracy-table11}
    \end{subfigure}
    \hfill
    \begin{subfigure}{0.18\textwidth}  
        \centering
        \raisebox{0.5mm}[0pt][0pt]{
      \begin{tikzpicture}
            \begin{axis}[
                ybar,
                ymin=0, ymax=1,
                ytick distance=0.2,
                enlargelimits=0.16,
                 width=1.3\textwidth,
                height=1.3\textwidth, 
                ylabel style={font=\fontsize{10}{10}\selectfont},
                yticklabel style={font=\fontsize{6}{6}\selectfont},
                symbolic x coords={RF, DT, KNN},
                xtick=data,
                xticklabel style={font=\fontsize{7}{7}\selectfont},
                bar width=3pt,  
                grid=both,
                grid style={line width=0.1pt, draw=gray!10},
                major grid style={line width=0.1pt,draw=black!30, dashed},
            ]
                \addplot[ybar, fill=blue!50!white, draw=blue!50!black] coordinates {(RF,0.99) (DT,0.99) (KNN,0.90)};
                \addplot[ybar, fill=red!50!white, draw=red!50!black] coordinates {(RF,0.95) (DT,0.91) (KNN,0.90)};
            \end{axis}
        \end{tikzpicture}
        }
        \caption{Model accuracy: Untargeted attack effect. \autoref{fig:Model-Metrics-table12} references Precision, Recall, and F1-Score metrics.}   \vspace{3.5mm}
        \label{fig:Model-Accuracy-table12}
    \end{subfigure}
    \hfill
    \begin{subfigure}{0.18\textwidth}  
        \centering
        \begin{tikzpicture}
            \begin{axis}[
                ybar,
                ymin=0, ymax=1,
                ytick distance=0.2,
                enlargelimits=0.16,
                width=1.3\textwidth,
                height=1.3\textwidth,  
                ylabel style={font=\fontsize{10}{10}\selectfont},
                yticklabel style={font=\fontsize{6}{6}\selectfont},
                symbolic x coords={RF, DT, KNN},
                xtick=data,
                xticklabel style={font=\fontsize{7}{7}\selectfont},
                bar width=3pt,  
                grid=both,
                grid style={line width=0.1pt, draw=gray!10},
                major grid style={line width=.1pt,draw=black!30, dashed},
            ]
                \addplot[ybar, fill=blue!50!white, draw=blue!50!black] coordinates {(RF,0.97) (DT,0.97) (KNN,0.97)};
                \addplot[ybar, fill=red!50!white, draw=red!50!black] coordinates {(RF,0.80) (DT,0.80) (KNN,0.80)};
            \end{axis}
        \end{tikzpicture}
       \caption{Model accuracy: address features manipulation effect.\autoref{fig:Table-13} references Precision, Recall, and F1-Score metrics. }
        \label{fig:Model-Accuracy-table13}
    \end{subfigure}

    \caption{Combined model accuracy across 2-6 figures.}
    \label{fig:Combined-Model-Accuracy}
\end{figure*}
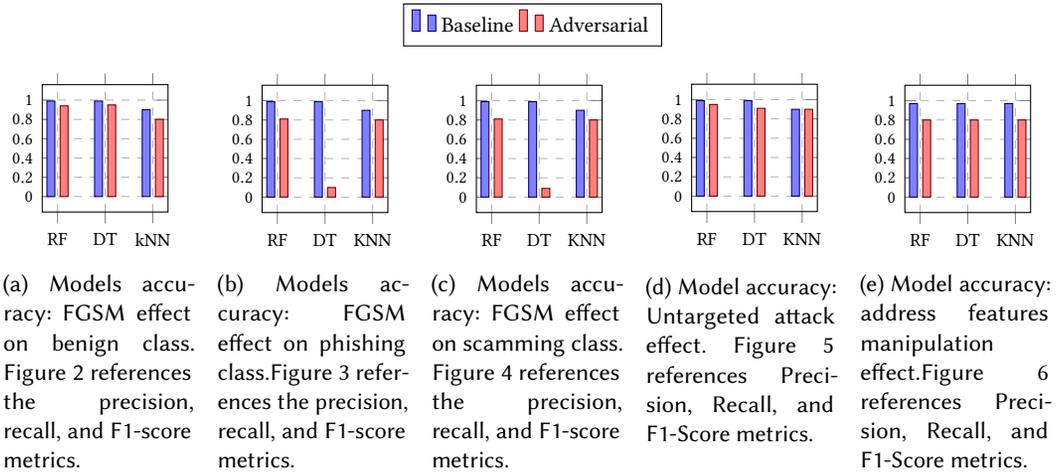

\begin{table*}[ht]
\centering
\caption{Impact of unseen address features on model performance for different models (RF, DT, KNN) using original and modified data. Metrics include accuracy, precision, recall, and F1-score for benign, phishing, and scamming classes.}
\label{table:unseen_address_D2} \vspace{-2mm}
\scalebox{0.75}{
\begin{tabular}{l|c|c|ccc|ccc|ccc}
\hline
\multirow{2}{*}{\textbf{Model}} & \multirow{2}{*}{\textbf{Data}} & \multirow{2}{*}{\textbf{Accuracy}} & \multicolumn{3}{c|}{\textbf{Precision}} & \multicolumn{3}{c|}{\textbf{Recall}} & \multicolumn{3}{c}{\textbf{F1-score}} \\
\cline{4-12}
 &  &  & \textbf{Benign} & \textbf{Phishing} & \textbf{Scam} & \textbf{Benign} & \textbf{Phishing} & \textbf{Scam} & \textbf{Benign} & \textbf{Phishing} & \textbf{Scam} \\ 
\hline

\multirow{2}{*}{RF}  
    & Original   & 0.97  & 0.98  & 0.95  & 0.99  & 1.00  & 0.74  & 0.93  & 0.99  & 0.83  & 0.96  \\  
    & Modified   & 0.80  & 0.80  & 0.00  & 0.00  & 1.00  & 0.00  & 0.00  & 0.89  & 0.00  & 0.00  \\  
\hline

\multirow{2}{*}{DT}  
    & Original  & 0.97  & 0.97  & 0.94  & 0.99  & 1.00  & 0.74  & 0.93  & 0.99  & 0.83  & 0.96  \\  
    & Modified  & 0.80  & 0.80  & 0.00  & 0.00  & 1.00  & 0.00  & 0.00  & 0.89  & 0.00  & 0.00  \\  
\hline

\multirow{2}{*}{KNN)}  
    & Original & 0.97  & 0.98  & 0.88  & 0.97  & 1.00  & 0.76  & 0.93  & 0.99  & 0.81  & 0.95  \\  
    & Modified  & 0.80  & 0.80  & 0.00  & 0.00  & 1.00  & 0.00  & 0.00  & 0.89  & 0.00  & 0.00  \\  
\hline
\end{tabular}}
\vspace{-3mm}
\end{table*}





{\noindent \ding{174} \em Financial Features}
AEs targeting financial features ({\em value}, {\em gas}, {\em gas\_price}) led to a reduced RF's accuracy of 0.79. Similarly, the DT's accuracy dropped to 0.79. The KNN model’s accuracy slightly decreased to 0.90. The impact of these adversarial manipulations is particularly evident in the phishing precision and recall metrics. For example, under value manipulation, RF's phishing precision dropped from 0.69 to 0.24 and recall from 0.56 to 0.27. Similarly, DT's phishing precision decreased from 0.59 to 0.20, with recall dropping from 0.59 to 0.31. KNN also showed a reduction in phishing precision from 0.73 to 0.58 and recall from 0.51 to 0.39 under the same conditions. These results are detailed in \autoref{table:financial_D2}.

\begin{table*}[t]
\centering
\caption{Impact of financial feature manipulation on model performance for different models (RF, DT, KNN) using original and manipulated data. Metrics include accuracy, precision, recall, and F1-score for benign, phishing, and scamming classes.}
\label{table:financial_D2} \vspace{-4mm}
\scalebox{0.75}{
\begin{tabular}{l|c|c|ccc|ccc|ccc}
\hline
\multirow{2}{*}{\textbf{Model}} & \multirow{2}{*}{\textbf{Strategy}} & \multirow{2}{*}{\textbf{Accuracy}} & \multicolumn{3}{c|}{\textbf{Precision}} & \multicolumn{3}{c|}{\textbf{Recall}} & \multicolumn{3}{c}{\textbf{F1-score}} \\
\cline{4-12}
 &  &  & \textbf{Benign} & \textbf{Phishing} & \textbf{Scam} & \textbf{Benign} & \textbf{Phishing} & \textbf{Scam} & \textbf{Benign} & \textbf{Phishing} & \textbf{Scam} \\ 
\hline

\multirow{5}{*}{RF}  
    & Original   & 0.92  & 0.95  & 0.69  & 0.85  & 0.97  & 0.56  & 0.79  & 0.96  & 0.62  & 0.82  \\  
    & Value   & 0.79  & 0.90  & 0.24  & 0.53  & 0.87  & 0.27  & 0.60  & 0.88  & 0.26  & 0.56  \\  
    & Gas   & 0.90  & 0.93  & 0.69  & 0.83  & 0.97  & 0.45  & 0.73  & 0.95  & 0.54  & 0.78  \\  
    & Gas Price  & 0.91  & 0.92  & 0.74  & 0.87  & 0.98  & 0.48  & 0.70  & 0.95  & 0.58  & 0.78  \\  
    & Combined  & 0.78  & 0.87  & 0.25  & 0.48  & 0.88  & 0.19  & 0.48  & 0.87  & 0.21  & 0.48  \\  
\hline

\multirow{5}{*}{DT}  
    & Original   & 0.91  & 0.95  & 0.59  & 0.81  & 0.95  & 0.59  & 0.80  & 0.95  & 0.59  & 0.80  \\  
    & Value   & 0.79  & 0.92  & 0.20  & 0.53  & 0.85  & 0.31  & 0.66  & 0.88  & 0.24  & 0.59  \\  
    & Gas   & 0.84  & 0.94  & 0.42  & 0.60  & 0.88  & 0.54  & 0.76  & 0.91  & 0.47  & 0.67  \\  
    & Gas Price  & 0.88  & 0.93  & 0.52  & 0.75  & 0.94  & 0.51  & 0.70  & 0.93  & 0.51  & 0.72  \\  
    & Combined  & 0.71  & 0.89  & 0.15  & 0.39  & 0.77  & 0.29  & 0.54  & 0.83  & 0.20  & 0.45  \\  
\hline

\multirow{5}{*}{KNN}  
    & Original   & 0.92  & 0.94  & 0.73  & 0.85  & 0.97  & 0.51  & 0.76  & 0.96  & 0.60  & 0.80  \\  
    & Value   & 0.90  & 0.93  & 0.58  & 0.80  & 0.96  & 0.39  & 0.74  & 0.94  & 0.46  & 0.77  \\  
    & Gas   & 0.89  & 0.93  & 0.68  & 0.74  & 0.95  & 0.47  & 0.72  & 0.94  & 0.56  & 0.73  \\  
    & Gas Price  & 0.89  & 0.92  & 0.67  & 0.80  & 0.96  & 0.43  & 0.68  & 0.94  & 0.52  & 0.73  \\  
    & Combined  & 0.86  & 0.91  & 0.53  & 0.67  & 0.94  & 0.31  & 0.64  & 0.92  & 0.39  & 0.65  \\  
\hline
\end{tabular}}
\vspace{-3mm}
\end{table*}

\begin{table*}[t]
\centering
\caption{Impact of temporal feature manipulation on model performance for different models (RF, DT, KNN) using original and manipulated data. Metrics include accuracy, precision, recall, and F1-score for benign, phishing, and scamming classes.}
\label{table:Block-D2} \vspace{-3mm}
\scalebox{0.75}{
\begin{tabular}{l|c|c|ccc|ccc|ccc}
\hline
\multirow{2}{*}{\textbf{Model}} & \multirow{2}{*}{\textbf{Strategy}} & \multirow{2}{*}{\textbf{Accuracy}} & \multicolumn{3}{c|}{\textbf{Precision}} & \multicolumn{3}{c|}{\textbf{Recall}} & \multicolumn{3}{c}{\textbf{F1-score}} \\
\cline{4-12}
 &  &  & \textbf{Benign} & \textbf{Phishing} & \textbf{Scam} & \textbf{Benign} & \textbf{Phishing} & \textbf{Scam} & \textbf{Benign} & \textbf{Phishing} & \textbf{Scam} \\ 
\hline

\multirow{4}{*}{RF}  
    & Original   & 0.99  & 1.00  & 0.88  & 0.97  & 1.00  & 0.91  & 0.96  & 1.00  & 0.89  & 0.97  \\  
    & Timestamp   & 0.80  & 0.80  & 1.00  & 1.00  & 1.00  & 0.00  & 0.00  & 0.89  & 0.01  & 0.00  \\  
    & Block Number   & 0.95  & 1.00  & 0.46  & 0.80  & 1.00  & 0.19  & 0.94  & 1.00  & 0.27  & 0.86  \\  
    & Combined   & 0.80  & 0.80  & 0.00  & 0.89  & 1.00  & 0.00  & 0.01  & 0.89  & 0.00  & 0.01  \\  
\hline

\multirow{4}{*}{DT}  
    & Original   & 0.99  & 1.00  & 0.87  & 0.97  & 1.00  & 0.91  & 0.96  & 1.00  & 0.89  & 0.97  \\  
    & Timestamp   & 0.80  & 0.80  & 1.00  & 1.00  & 1.00  & 0.00  & 0.00  & 0.89  & 0.01  & 0.00  \\  
    & Block Number   & 0.95  & 1.00  & 0.46  & 0.80  & 1.00  & 0.19  & 0.94  & 1.00  & 0.27  & 0.86  \\  
    & Combined   & 0.80  & 0.80  & 0.00  & 0.89  & 1.00  & 0.00  & 0.01  & 0.89  & 0.00  & 0.01  \\  
\hline

\multirow{4}{*}{KNN}  
    & Original   & 0.99  & 1.00  & 0.92  & 0.96  & 1.00  & 0.87  & 0.98  & 1.00  & 0.89  & 0.97  \\  
    & Timestamp   & 0.80  & 0.80  & 1.00  & 1.00  & 1.00  & 0.00  & 0.00  & 0.89  & 0.01  & 0.00  \\  
    & Block Number   & 0.94  & 1.00  & 0.52  & 0.77  & 0.99  & 0.20  & 0.95  & 1.00  & 0.29  & 0.85  \\  
    & Combined   & 0.80  & 0.80  & 0.00  & 0.85  & 1.00  & 0.00  & 0.01  & 0.89  & 0.00  & 0.02  \\  
\hline
\end{tabular}}
\vspace{-3mm}
\end{table*}

{\noindent \ding{175} \em Using Temporal Features}
Adversarial manipulations of temporal features ({\em block\_timestamp}, {\em block\_number}) showed the RF model's accuracy fell from 0.99 to 0.80, with phishing detection metrics nearly nullified. This significant drop indicates that temporal features are a major vulnerability for RF and likely for other models, which adversaries can exploit to severely disrupt the model's ability to detect malicious activities. For instance, under timestamp manipulation, RF's phishing recall dropped drastically from 0.91 to zero. The DT’s accuracy similarly declined to 0.80. The KNN model's accuracy also dropped to 80.26\%. These results are further detailed in \autoref{table:Block-D2}, which compares model performance across different strategies, including original, timestamp-manipulated, block number-manipulated, and combined approaches, highlighting the models' vulnerabilities to temporal feature perturbations.

\takeaway{These results underscore the need for robust defensive mechanisms against AEs. The significant decline in performance metrics under simple conditions highlight the need for a more reliable classification of transactions.}

\section{Defense Mechanism} \label{sec:defense}
In this section, we propose a defense mechanism to mitigate the impact of adversarial attacks on Ethereum phishing transaction detection models. Given the vulnerability of machine learning models to adversarial perturbations, our approach centers around adversarial training~\cite{LiZRC23,ZhangJCH24}, which enhances model robustness by incorporating AEs into the training process and allowing the model to learn to resist manipulations. We note that it is important to distinguish adversarial training from traditional data poisoning. While data poisoning aims to compromise model performance by injecting misleading data, adversarial training is a defensive measure designed to improve model resilience to real-world adversarial scenarios.

\subsection{Adversarial Training Approach}

We applied adversarial training to defend against targeted adversarial attacks across all classes—benign, phishing, and scamming—which were individually targeted. We also focused on specific feature manipulations, particularly the \textit{TimeStamp} and \textit{Value} features. To implement the adversarial training approach, we followed these steps.

\BfPara{\ding{172} Generating Adversarial Examples} We employed existing adversarial example generation functions to create targeted samples by manipulating key features. This involved generating AEs with minimal perturbations to the \textit{TimeStamp} and \textit{Value} features. Moreover, we used FGSM to craft targeted AEs by perturbing the input features associated with benign, phishing, and scamming labels, aiming to challenge the models with realistic adversarial scenarios~\cite{ChenQDYT20}. These perturbations were carefully controlled to remain within realistic bounds, ensuring that the examples would simulate genuine attack scenarios while minimizing potential performance compromises.

\BfPara{\ding{173} Models Retraining}The RF, DT, and KNN models were retrained using an augmented dataset that included both original and adversarially modified data. During this process, model parameters were adjusted to improve resilience against adversarial attacks. Unlike poisoning, which reduces model performance, the AEs used in this study were intended to identify weaknesses in key feature spaces that influence classification accuracy. This method aimed to help the models generalize to new attacks without becoming overly adapted to AEs. By introducing potentially misleading patterns, the goal was to enhance the models’ ability to classify transactions, particularly those in categories such as benign and phishing.

\BfPara{\ding{174} Models Evaluation} After training with AEs, the models were tested on both the original and modified sets to evaluate their ability to recognize manipulated. The assessment focused on key performance metrics, including accuracy, precision, recall, and F1 score, while also examining misclassification patterns to understand how the models responded to adversarial attacks.

\section{Post Retraining Results}
\subsection{Preliminary Results}
After applying timestamp and value manipulations, we assess the impact of adversarial retraining on the RF, DT, and KNN models. We analyze how retraining enhances model accuracy and resilience, focusing on restoring precision, recall, and F1 scores across all models.

\BfPara{Timestamp} Before retraining, timestamp manipulations caused significant accuracy reductions, with RF at 0.95, DT at 0.94, and KNN at 0.83 (Table \ref{table:time_minap}). After retraining, these models showed enhanced resilience, maintaining accuracies of 0.98 for RF and DT, and improving to 0.94 for KNN (Table \ref{table:TMretrained}). Stability in precision, recall, and F1 scores across all models indicates that adversarial training effectively mitigated the adverse effects of timestamp manipulations. The differential impact on KNN can be attributed to its inherent sensitivity to the input space distribution, which makes timestamp manipulations disruptive. We note that while we observed slight variances in the non-adversarial test sets after retraining, these were negligible compared to the significant gains in adversarial robustness.

\begin{table*}[t]
\centering
\caption{Performance of RF, DT, and KNN after adversarial training on timestamp manipulations for different models, increments, and datasets. Metrics include accuracy, precision, recall, F1-score, and instance count for benign and phishing classes.}
\label{table:TMretrained}\label{table:time_minap} \vspace{-5mm}
\scalebox{0.75}{
\begin{tabular}{l|c|c|c|cc|cc|cc|cc}
\hline
\multirow{2}{*}{\textbf{Increment}} & \multirow{2}{*}{\textbf{Model}} & \multirow{2}{*}{\textbf{Dataset}} & \multirow{2}{*}{\textbf{Accuracy}} & \multicolumn{2}{c|}{\textbf{Precision}} & \multicolumn{2}{c|}{\textbf{Recall}} & \multicolumn{2}{c|}{\textbf{F1-score}} & \multicolumn{2}{c}{\textbf{Count}} \\
\cline{5-12}
 &  &  &  & \textbf{Benign} & \textbf{Phish} & \textbf{Benign} & \textbf{Phish} & \textbf{Benign} & \textbf{Phish} & \textbf{Benign} & \textbf{Phish} \\ 
\hline

\multirow{3}{*}{Original}  
    & RF  & Baseline  & 0.98 & 1.00 & 0.97 & 0.98 & 1.00 & 0.99 & 0.98 & 15,989 & 7,483 \\  
    & DT  & Baseline  & 0.98 & 1.00 & 0.95 & 0.98 & 1.00 & 0.99 & 0.97 & 15,989 & 7,483 \\  
    & KNN & Baseline  & 0.95 & 1.00 & 0.88 & 0.94 & 1.00 & 0.97 & 0.94 & 15,989 & 7,483 \\  
\hline

\multirow{3}{*}{+24 Hours}  
    & RF  & Adversarial & 0.98 & 1.00 & 0.97 & 0.98 & 1.00 & 0.99 & 0.98 & 15,989 & 7,483 \\  
    & DT  & Adversarial & 0.98 & 1.00 & 0.95 & 0.98 & 1.00 & 0.99 & 0.97 & 15,989 & 7,483 \\  
    & KNN & Adversarial & 0.94 & 1.00 & 0.86 & 0.92 & 1.00 & 0.96 & 0.92 & 15,989 & 7,483 \\  
\hline

\multirow{3}{*}{+1 Hour}  
    & RF  & Adversarial & 0.98 & 1.00 & 0.97 & 0.98 & 1.00 & 0.99 & 0.98 & 15,989 & 7,483 \\  
    & DT  & Adversarial & 0.98 & 1.00 & 0.95 & 0.98 & 1.00 & 0.99 & 0.97 & 15,989 & 7,483 \\  
    & KNN & Adversarial & 0.95 & 1.00 & 0.88 & 0.94 & 1.00 & 0.97 & 0.94 & 15,989 & 7,483 \\  
\hline

\multirow{3}{*}{+30 Minutes}  
    & RF  & Adversarial & 0.98 & 1.00 & 0.97 & 0.98 & 1.00 & 0.99 & 0.98 & 15,989 & 7,483 \\  
    & DT  & Adversarial & 0.98 & 1.00 & 0.95 & 0.98 & 1.00 & 0.99 & 0.97 & 15,989 & 7,483 \\  
    & KNN & Adversarial & 0.95 & 1.00 & 0.88 & 0.94 & 1.00 & 0.97 & 0.94 & 15,989 & 7,483 \\  
\hline

\multirow{3}{*}{+15 Minutes}  
    & RF  & Adversarial & 0.98 & 1.00 & 0.97 & 0.98 & 1.00 & 0.99 & 0.98 & 15,989 & 7,483 \\  
    & DT  & Adversarial & 0.98 & 1.00 & 0.95 & 0.98 & 1.00 & 0.99 & 0.97 & 15,989 & 7,483 \\  
    & KNN & Adversarial & 0.95 & 1.00 & 0.89 & 0.94 & 1.00 & 0.97 & 0.94 & 15,989 & 7,483 \\  
\hline

\multirow{3}{*}{+5 Minutes}  
    & RF  & Adversarial & 0.98 & 1.00 & 0.97 & 0.98 & 1.00 & 0.99 & 0.98 & 15,989 & 7,483 \\  
    & DT  & Adversarial & 0.98 & 1.00 & 0.95 & 0.98 & 1.00 & 0.99 & 0.97 & 15,989 & 7,483 \\  
    & KNN & Adversarial & 0.95 & 1.00 & 0.88 & 0.94 & 1.00 & 0.97 & 0.94 & 15,989 & 7,483 \\  
\hline

\end{tabular}}
\vspace{-3mm}
\end{table*}

\BfPara{Value} Value manipulations significantly affected the RF and DT models, reducing their accuracy from 0.99 and 0.98 to 0.69, respectively (Table \ref{table:value_manipulations}). After adversarial retraining (Table \ref{table:VMretrained}), both models regained their original accuracy, while KNN remained stable. This suggests that adversarial training helped restore classification performance under value manipulations. The differences in how RF and DT responded to these changes reflect variations in their handling of feature splits and decision thresholds, with DT being more sensitive to abrupt shifts in feature values than RF.

\begin{table*}[t]
\centering
\caption{Performance evaluation of RF, DT, and KNN models after adversarial training on value manipulation, subjected to 1\% uniform (U) and proportional (P) value manipulation strategies compared to the original (O). Metrics include accuracy, precision, recall, F1-score, and instance count for benign and phishing classes.}
\label{table:VMretrained} \vspace{-2mm}
\scalebox{0.75}{
\begin{tabular}{l|c|c|c|cc|cc|cc|cc}
\hline
\multirow{2}{*}{\textbf{Model}} & \multirow{2}{*}{\textbf{Strategy}} & \multirow{2}{*}{\textbf{Accuracy}} & \multicolumn{2}{c|}{\textbf{Precision}} & \multicolumn{2}{c|}{\textbf{Recall}} & \multicolumn{2}{c|}{\textbf{F1-score}} & \multicolumn{2}{c}{\textbf{Count}} \\
\cline{4-11}
 &  &  & \textbf{Benign} & \textbf{Phish} & \textbf{Benign} & \textbf{Phish} & \textbf{Benign} & \textbf{Phish} & \textbf{Benign} & \textbf{Phish} \\ 
\hline

\multirow{3}{*}{RF}  
    & Original   & 0.99  & 0.98  & 1.00  & 1.00  & 0.99  & 0.99  & 0.99  & 7,483  & 15,989  \\  
    & Uniform   & 0.99  & 0.98  & 1.00  & 1.00  & 0.99  & 0.99  & 0.99  & 7,661  & 15,811  \\  
    & Proportional & 0.99  & 0.98  & 1.00  & 1.00  & 0.99  & 0.99  & 0.99  & 7,652  & 15,820  \\  
\hline

\multirow{3}{*}{DT}  
    & Original  & 0.98  & 0.95  & 1.00  & 1.00  & 0.97  & 0.97  & 0.99  & 7,483  & 15,989  \\  
    & Uniform  & 0.98  & 0.95  & 1.00  & 1.00  & 0.97  & 0.97  & 0.99  & 7,888  & 15,584  \\  
    & Proportional  & 0.98  & 0.95  & 1.00  & 1.00  & 0.97  & 0.97  & 0.99  & 7,884  & 15,588  \\  
\hline

\multirow{3}{*}{KNN}  
    & Original  & 0.98  & 0.93  & 1.00  & 1.00  & 0.96  & 0.96  & 0.98  & 8,064  & 15,408  \\  
    & Uniform  & 0.98  & 0.93  & 1.00  & 1.00  & 0.96  & 0.96  & 0.98  & 8,065  & 15,407  \\  
    & Proportional  & 0.98  & 0.93  & 1.00  & 1.00  & 0.96  & 0.96  & 0.98  & 8,064  & 15,408  \\  
\hline

\end{tabular}}
\vspace{-3mm}
\end{table*}

\subsection{Post Retraining Results of Targeted Attacks}
After retraining, the RF, DT, and KNN models were tested to see how they responded to targeted attacks on different classes. The results showed improvements in their ability to correctly classify benign, phishing, and scamming cases. The KNN model had a slightly higher number of misclassifications, likely because it relies on local neighborhood data, which can be less reliable when class boundaries are altered. Despite this, the overall findings show that adversarial training helped improve model performance. 

\BfPara{Benign Class} Benign class accuracy was severely impacted by targeted adversarial attacks, with RF and DT accuracies dropping from 1.00 and 0.99 to 0.84, and KNN from 0.97 to 0.90—indicating significant performance declines of over 15\% and 7\%, respectively (Table \ref{table:Benign-target}). These attacks caused notable misclassifications, especially in the KNN model, where 60 benign instances were incorrectly classified as phishing and 97 as scamming. However, after adversarial training, the models showed substantial recovery (Table\ref{table:BenignRetrained}). Both RF and DT regained their high accuracy levels of 0.99, and KNN improved to 0.98, reflecting a strong restoration of performance. Misclassifications were drastically reduced, with RF and DT eliminating nearly all errors, while KNN minimized its misclassification rates to just 0.22\% for phishing and 0.03\% for scamming. Overall, adversarial training proved highly effective in restoring and even enhancing the robustness of these models against adversarial attacks, particularly for the benign class, though KNN still showed slight residual vulnerabilities.

\BfPara{Phishing Class} Phishing class accuracy was drastically affected by targeted attacks, with RF and DT dropping to 0.01 and KNN to 0.02—reflecting a more than 95\% decrease in performance (Table \ref{table:Phishing-target}). However, after adversarial training, these models showed significant resilience. Both RF and DT rebounded to an accuracy of 0.99, and KNN improved to 0.98, indicating a near-complete recovery (Table \ref{table:PhishingRetrained}). Misclassifications were nearly eliminated, with RF achieving 0\% errors, and DT showing minimal errors. KNN, while much improved, still displayed some vulnerability with small but noticeable misclassification rates. Overall, adversarial training has proven highly effective, particularly for RF and DT, though KNN may benefit from further refinement to achieve similar levels of robustness.

\begin{table*}[t]
\centering
\caption{Impact of adversarial training using FGSM on RF, DT, and KNN models for benign, phishing, and scamming classes. The table presents accuracy before (Pre) and after (Post) retraining, as well as misclassification rates for different instance types. Cells of repeated values are merged.}
\label{table:BenignRetrained} \label{table:PhishingRetrained} \label{table:ScammingRetrained}\vspace{-2mm}
\scalebox{0.85}{
\begin{tabular}{l|cc|cc|cc}
\hline
\textbf{Metric} & \multicolumn{2}{c|}{\textbf{Random Forest (RF)}} & \multicolumn{2}{c|}{\textbf{Decision Tree (DT)}} & \multicolumn{2}{c}{\textbf{K-Nearest Neighbors (KNN)}} \\  
\hline
 & \textbf{Pre} & \textbf{Post} & \textbf{Pre} & \textbf{Post} & \textbf{Pre} & \textbf{Post} \\  
\hline
\multicolumn{7}{c}{\textbf{Accuracy Before and After Adversarial Training}} \\  
\hline
Benign Detection   &  \multicolumn{4}{c|}{0.99} &  \multicolumn{2}{c}{0.98} \\  \hline
Phishing Detection &  \multicolumn{4}{c|}{0.99} & \multicolumn{2}{c}{0.98} \\  \hline
Scamming Detection & \multicolumn{4}{c|}{0.99} & \multicolumn{2}{c}{0.98} \\  
\hline
\multicolumn{7}{c}{\textbf{Instance Counts Before Adversarial Training}} \\  
\hline
\multicolumn{1}{c|}{\bf Benign} & \multicolumn{2}{c|}{\bf Phishing} & \multicolumn{2}{c|}{\bf Scamming} & \multicolumn{2}{c}{\bf Fake ICO} \\\hline
\multicolumn{1}{c|}{11,431} & \multicolumn{2}{c|}{629} & \multicolumn{2}{c|}{2,189} & \multicolumn{2}{c}{2,189}\\
\hline
\multicolumn{7}{c}{\textbf{Misclassification Rates After Adversarial Training}} \\  
\hline
\multicolumn{7}{c}{\textbf{Benign Misclassifications}}  \\ \hline 
Benign → Phishing      & \multicolumn{5}{c|}{0.00\%} & 0.22\% \\  \hline 
Benign → Scamming      & \multicolumn{5}{c|}{0.00\%}& 0.03\% \\  \hline
Phishing → Benign      & \multicolumn{5}{c|}{0.00\%}& 8.43\% \\ \hline 
Scamming → Benign      & \multicolumn{3}{c|}{0.00\%}& \multicolumn{1}{c|}{0.14\%\%} & \multicolumn{1}{c|}{0.00\%} & 2.69\% \\  
\hline
\multicolumn{7}{c}{\textbf{Phishing Misclassifications}}  \\ \hline 
Phishing → Benign      & \multicolumn{2}{c|}{0.00\%} & \multicolumn{2}{c|}{1.75\%}  & \multicolumn{2}{c}{9.70\%} \\  \hline
Phishing → Scamming    & \multicolumn{4}{c|}{0.00\%} & \multicolumn{2}{c}{8.11\%} \\  \hline
Scamming → Phishing    & \multicolumn{2}{c|}{0.32\%} & \multicolumn{2}{c|}{0.55\%} & \multicolumn{2}{c}{0.73\%} \\  
\hline
\multicolumn{7}{c}{\textbf{Scamming Misclassifications}}  \\ \hline 
Scamming → Benign      & \multicolumn{4}{c|}{0.00\%} & \multicolumn{2}{c}{2.69\%} \\  \hline
Scamming → Phishing    & \multicolumn{4}{c|}{0.00\%} & \multicolumn{2}{c}{1.14\%} \\  \hline
Phishing → Scamming    & \multicolumn{2}{c|}{4.77\%} & \multicolumn{2}{c|}{6.20\%} & \multicolumn{2}{c}{7.79\%} \\ \hline 
Fake ICO → Scamming    & \multicolumn{4}{c|}{0.00\%} & \multicolumn{2}{c}{100\%} \\  
\hline
\end{tabular}}
\vspace{-5mm}
\end{table*}

\BfPara{Scamming Class} Scamming class accuracy was significantly impacted by targeted adversarial attacks, with RF and DT accuracies dropping from 0.99 and 0.98 to 0.14, and KNN plummeting from 0.67 to 0.07—representing drops of over 85\% and 90\%, respectively (Table \ref{table:Scamming-target}). These attacks led to a substantial number of scamming instances being misclassified, particularly in the KNN model, where 1,214 instances were incorrectly classified as benign. However, after applying adversarial training, the models demonstrated significant recovery (Table \ref{table:ScammingRetrained}). RF and DT accuracies returned to 0.99, and KNN improved to 0.98, showing a strong restoration of performance. Misclassifications were greatly reduced, with RF achieving a 0\% misclassification rate for scamming instances, and DT and KNN reducing errors to minimal levels. Despite these improvements, KNN still showed some residual vulnerability, with 2.69\% of scamming instances misclassified as benign and 1.14\% as phishing. Overall, adversarial training proved highly effective for most models, particularly RF and DT, though KNN may still benefit from further refinement to achieve comparable robustness.


\section{Discussion}

\BfPara{Feature Selection for Optimal Classification} This study examines how feature selection affects ML models in transaction classification. The choice of features impacts a model’s ability to classify transactions and resist manipulation. Among the tested features, \textbf{timestamp} and \textbf{value} impacted model performance. Their stability under manipulation suggests they capture transaction details that differentiate legitimate from fraudulent activities. Timestamp changes affected all models, with RF handling them better than DT and KNN. This indicates that time-related features contribute to fraud detection, as transaction timing can reveal patterns linked to fraudulent behavior. The accuracy results show that transaction time and date help identify fraud. 

\textbf{Value} manipulations, including uniform and proportional changes, affected model performance, especially under uniform. RF and DT models showed accuracy drops, while KNN remained stable. After adversarial training, RF and DT regained accuracy, showing that adversarial training helps reduce the impact of value changes. The results indicate that transaction value is an important feature in classification. The ability to manipulate transaction value without detection suggests a potential risk, as it may enable bypassing of fraud detection systems.

\BfPara{Most Resistant Features to Adversarial Attacks} The results show that \textbf{address features}, specifically the \texttt{From} and \texttt{To} addresses, are less affected by adversarial attacks than other features. This may be because address features capture transaction patterns that are harder to change. When these features were manipulated, DT and RF models showed some accuracy reduction, while KNN was more affected. The differences across models suggest that how addresses are represented and how each model processes them influence their resistance to attacks. Address features may hold relationship patterns between transactions that remain stable even when altered.

In addition, the analysis showed that while important for accuracy, temporal features were also resistant to manipulations. This suggests that temporal data capture transaction patterns that are harder to alter without making significant changes. Shifting timestamps led to accuracy declines, but the impact was smaller than value manipulations. This indicates that temporal features play a key role in classification and are less vulnerable to adversarial attacks due to the complexity and uniqueness of timestamp data.

Adversarial training is an effective way to improve model robustness against targeted attacks. Exposing models to AEs during training made them more resistant to critical features such as \textbf{timestamp} and \textbf{value} disruptions. This approach helped compare the effects of such attacks, showing that adversarial training can be a practical method for reinforcing phishing detection models against exploitable weaknesses.

\BfPara{Best Combinations} For a more resilient classification, combining \textbf{temporal and address features} has proven to be effective. This integration takes advantage of both temporal patterns and relational data, creating models that  less vulnerable to adversarial interference. These features provide a layer of protection. Temporal features capture transaction timing and distribution patterns, while address features offer a stable relational structure that is harder to manipulate. This approach strengthens the model’s accuracy, even when adversarial attacks attempt to exploit either feature. 

Similarly, combining\textbf{temporal features with financial features}, such as transaction value and gas price, further enhances robustness. Although financial features can be uniformly manipulated, combining them with temporal data provides additional context that improves the model’s resilience. The temporal features help to contextualize the financial data, mitigating the impact of adversarial value manipulations.

The following recommendations can be drawn for the effective and robust classification of transactions, especially in adversarial attacks. 
    \ding{172} \textbf{Focus on Temporal and Address Features}: Timestamp and address data could be key features in classification models, as they offer resilience against adversarial manipulations and play a crucial role in maintaining accuracy.
    \ding{173} \textbf{Integrate Financial Features with Temporal Data}: Use financial transaction data with temporal features to improve robustness and provide a transactional context that helps counteract adversarial manipulations.
    \ding{174} \textbf{Adopt a Multi-Feature Approach}: Utilize a combination of diverse feature types to leverage their respective strengths and ensure a balanced, resilient classification model capable of withstanding various adversarial manipulations.

\section{Conclusion}\label{sec: Conclusion}

This study examines how machine learning models can be affected by adversarial attacks in detecting fraudulent transactions. The results show that different classifiers have varying levels of vulnerability. RF is more resistant, while DT and KNN are more easily influenced by these attacks. This highlights the need for careful feature selection and adversarial training to improve model robustness. These techniques help reduce the impact of attacks and improve classification accuracy. Using temporal and address features along with financial data strengthens model defenses and supports reliable classification under adversarial conditions. The adversarial training approach tested in this study also helps reduce the effects of targeted attacks, making it a valuable defense strategy. Future research will explore more complex attack scenarios and apply these methods to other financial platforms to evaluate their applicability and effectiveness.


\section*{Acknowledgement}
An earlier version of this work appeared in proceedings of WISA 2024 as ``Simple Perturbations Subvert Ethereum Phishing Transactions Detection''~\cite{AlghureidD24}.


\end{document}